\documentclass[fleqn,usenatbib]{mnras}

\usepackage{newtxtext,newtxmath}

\usepackage[T1]{fontenc}

\DeclareRobustCommand{\VAN}[3]{#2}
\let\VANthebibliography\thebibliography
\def\thebibliography{\DeclareRobustCommand{\VAN}[3]{##3}\VANthebibliography}
\usepackage[normalem]{ulem}
\usepackage{xcolor}
\definecolor{RoyalBlue}{rgb}{0.25,0.41,0.88}

\usepackage{graphicx}	
\usepackage{amsmath}	

\title[SHH method for SIDM]{An SIDM-hydro hybrid method for simulating self-interacting dark matter}

\author[S. Schon et al.]{Sarah Schon$^{1,2}$\thanks{E-mail: sqs7027@psu.edu},
Michael Ryan$^{3}$,
John Blakely$^{1,4}$,
Jeysen Flores-Velazquez$^{1,2}$,
\newauthor Sarah Shandera$^{1,2}$,
Donghui Jeong$^{1,4,5}$
\\
$^{1}$Institute for Gravitation and the Cosmos, The Pennsylvania State University, University Park, PA 16802, USA\\
$^{2}$Department of Physics, The Pennsylvania State University, University Park, PA, 16802, USA\\
$^{3}$Center for Cosmology, Department of Physics and Astronomy, University of California - Irvine, Irvine, CA 92697, USA\\
$^{4}$Department of Astronomy and Astrophysics, The Pennsylvania State University, University Park, PA, 16802, USA\\
$^{5}$School of Physics, Korea Institute for Advanced Study (KIAS), 85 Hoegiro, Dongdaemun-gu, Seoul, 02455, Republic of Korea
}

\date{Accepted XXX. Received YYY; in original form ZZZ}

\pubyear{\the\year{}}

\begin{document}
\label{firstpage}
\pagerange{\pageref{firstpage}--\pageref{lastpage}}
\maketitle

\begin{abstract}
We present a new scheme to couple existing numerical methods for elastic self-interacting dark matter (SIDM) to the hydrodynamic equations via a continuous function of the local Knudsen number. The method, an SIDM-hydro hybrid (SHH), allows more efficient simulation of the evolution of inhomogeneous halos deep into the regime of gravothermal collapse. With the improved efficiency gained by moving to a hydrodynamical description in high-density regions, the SHH method allows central densities of two orders of magnitude higher to be reached in considerably less simulation time than traditional methods. Our implementation should be considered as the first step toward a robust SHH method, as we interpolate the first and second moments of the Boltzmann equation in the ideal-fluid limit only. The simulation results are qualitatively similar to those found with other methods, although there are differences in the implementation of the primary physics driving the dynamics, and in the details of the resulting halo profiles. However, our results indicate that the SHH technique shows promise to investigate gravothermal collapse in diverse, dynamical environments. The method can be extended to incorporate non-ideal fluid terms and dissipation, as needed for dark-matter scenarios where interactions beyond the elastic regime may be important in the dense interiors of some halos. 
\end{abstract}



\section{Introduction}

Self-interacting dark matter (SIDM) models make up an intriguing allotment of the dark matter (DM) candidate landscape (see, e.g., \cite{Adhikari:2022sbh} for a recent review of motivation and constraints). While far from the only motivation, a historically prominent point of interest was the potential of even purely elastic interactions in SIDM to alleviate the ``cusp-core'' problem that arose due to an apparent tension between collisionless dark matter-only N-body simulations and galaxy density-profile observations \citep{de_Blok:1997,van_den_Bosch:2000,de_Blok:2001,de_Naray:2008,Oman:2015}. Although contemporary wisdom points to baryonic feedback \citep{McCarthy:2012,Sales:2016,DiCintio:2016,Santos-Santos:2016,Creasey:2017,Ferrero:2017,Katz:2017} as an important part of the solution to this particular problem, interest in self-interactions' impact on halo morphology has persisted \citep{Kamada:2017,Robertson:2018,Zavala:2019}. 

Understanding the effects of DM self-interactions on non-linear structure formation requires new numerical techniques. SIDM cannot be accurately treated either by N-body techniques for collisionless cold dark matter, nor by the same hydrodynamic codes that work for baryonic matter. The first proposed SIDM simulation algorithms added a small probability of elastic scattering to N-body particles \citep{Burkert:2000,Yoshida:2000,Dave:2001,Balberg_2002}. This treatment works well for the low-density outskirts of halos since observations constrain any DM self-interactions to be rather weak. However, self-interaction effects become more appreciable toward the higher-density core of a typical halo. Semi-analytic ``gravothermal fluid'' methods \citep{Balberg_2002,Koda:2011,Essig_2019,Yang:2022zkd,Yang2023,Outmezguine:2023,Gad-Nasr:2023gvf}, akin to those used for the gravothermal catastrophe studied in globular clusters \citep{Lynden-Bell1968, Lynden-Bell1980}, have been developed  to model how the heat transfer facilitated by SIDM collisions can alter the central regions of a halo. Recently, \cite{gurian:2025,kamionkowski:2025} developed one-dimensional particle based codes which are intermediate in computational cost between three-dimensional simulations and fluid models. These codes conserve energy more accurately than three-dimensional simulations yet do not require the equilibrium assumptions and calibration parameter(s) of the fluid models. 

The semi-analytic techniques are especially important since the dramatic increase in density during the gravothermal collapse of the core poses a number of numerical challenges. In order for the traditional SIDM numerical approach to be valid, the simulation must remain in the regime where the probability for scattering for each N-body particle remains small. As the density increases in the core region, this requires decreasing the timestep in the simulation, dramatically increasing the computational cost. In addition, the force softening must be treated with care in high-density regions, as multiple N-body particles can accumulate below the scale where the forces are resolved. Finally, noise from the stochastic implementation of the elastic scattering, where the angle is chosen at random \citep{Mace:2024uze}, is higher in denser regions, including within sub-structure. Recent systematic comparative studies of numerical techniques and semi-analytic results demonstrate the sensitivity of the quantitative results to choices in the simulation parameters \citep{Palubski:2024ibb,Fischer:2024eaz, Mace:2024uze}. More fundamentally, in high density regimes the mean free path for self-scattering can become very small compared to the other scales in the problem, the regime of small Knudsen number\footnote{Where the Knudsen number is taken to be the ratio of the particle's mean free path to a relevant characteristic length scale.}, and a fluid description is needed to efficiently and accurately describe the evolution. At least some of the issues observed in current simulations would be straight-forwardly alleviated if hydrodynamical techniques could be used in the regime of small Knudsen number. 

\begin{figure}
	\includegraphics[width=\columnwidth]{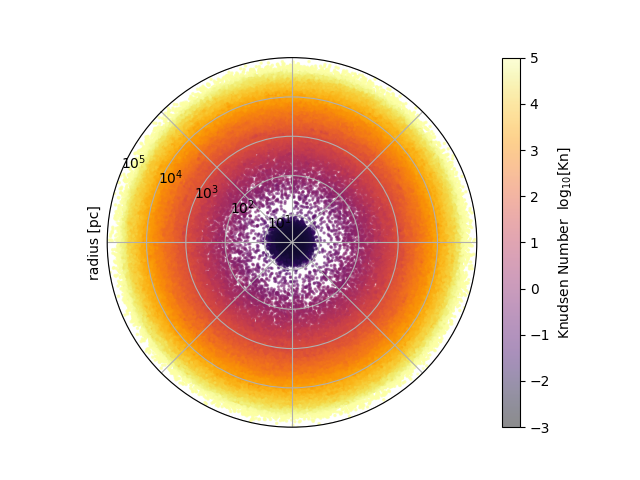}
    \caption{Polar plot of the the local Knudsen number of a gravothermally collapsed halo, calculated by comparing the mean free path given SIDM interactions and the local scale of dynamics for the simulation particles (Eq.(\ref{equ:Knudsen_sims})). The color bar indicates the Knudsen number in logarithmic scale $\log({\rm Kn})$. If the mean free path is much smaller than the dynamical scale, the fluid description, rather than N-body, is appropriate. \label{fig:Kn_polar}}
\end{figure}

In this work, we propose to efficiently extend numerical SIDM methods in the frequent scattering (high-density) limit by introducing a numerical buffer zone that allows smooth interpolation from the N-body plus elastic scattering algorithm, which is valid in regions with high Knudsen number, to a fluid treatment in regions with low Knudsen number. We use the publicly available simulation package GIZMO \citep{Hopkins2015} and the incorporated SIDM modules \citep{Rocha2013,Robles_2017} to evolve halos for which the local Knudsen number for the dark matter varies by as much as ten orders of magnitude between the outskirts and central region. Figure \ref{fig:Kn_polar} shows a representative example. In the next Section we review the current semi-analytic and numerical approaches to halo evolution with SIDM and introduce a new approach. Section \ref{sec:implementation} provides details of how the new method is implemented, and Section \ref{sec:sims} contains the results of our simulations. We summarize in Section \ref{sec:conclude}.

\section{An SIDM-hydro hybrid method}
\label{sec:Method}
In this section, we first review semi-analytic methods and benchmark their outputs against those from standard SIDM simulations. We then introduce a new interpolation technique—inspired by the semi-analytic approach but designed to work in any simulation without symmetry assumptions. Like the gravothermal collapse method, our scheme isn’t a fully first-principles solution; it relies on a single adjustable parameter that can be calibrated against independent data. Finally, we describe the form of this parameterization and demonstrate how it shapes the results.

\subsection{Review of the gravothermal, iso-thermal, and standard SIDM simulation methods}
\label{sec:SAreview}
Semi-analytic techniques for treating the high-density regime capture some effects of interactions at finite mean free path in self-gravitating systems without the full complexity of the Navier-Stokes equations \citep{Lynden-Bell1980,Balberg_2002, Koda:2011,Essig_2019,Yang2022, Yang2023,Outmezguine:2023,Gad-Nasr:2023gvf}. This approach dates back to \citet{Lynden-Bell1980}, who used dimensional analysis to propose effective thermodynamic equations valid even in the long mean free path limit. 

For SIDM, the notion of long or short mean free path, $\lambda_{\rm mfp}$, is determined by comparing the to the gravitational scale height, $H=\nu/\sqrt{4\pi G\rho(r)}$ where $\nu$ is the 1-D velocity dispersion and $\rho(r)$ is the radial density profile. That is, the short mean free path regime corresponds to small Knudsen number
\begin{equation}
\label{eq:KnGravothermal}
{\rm Kn_{gravothermal}}=\frac{\lambda_{\rm mfp}}{H}\ll1\,,
\end{equation}
and vice versa\footnote{The Knudsen number defined relative to the scale height here differs to that used in Figure \ref{fig:Kn_polar}, where $\mathrm{Kn}$ is taken relative to the numerical cell size. In general, both the physical and simulation Knudsen number need to be taken into consideration in a numerical context, though for the remainder of this work we restrict ourselves to the simulation Knudsen number.}. For SIDM, $\lambda_{\rm mfp}=\frac{1}{n\sigma}$ depends on the self-interaction cross-section $\sigma$ and the number density of particles, $n$. Then heat conduction, or dissipation of gravitational energy induced by particle self-scattering, at any local position or density $\rho(r)$ is estimated as an interpolation between conductivities, $\kappa$, in the large and small mean free path (smfp and lmfp) limits:
\begin{equation}
\kappa = (\kappa_{\rm smfp}^{-1} + \kappa_{\rm lmfp}^{-1})^{-1}\,.
\label{eq:kappa}
\end{equation}
The short mean free path conductivity, $\kappa_{\rm smfp}$, can be found analytically \citep{ChapmanCowlingBook, enskog1917} by performing an expansion of the fluid equations in small Knudsen number, away from the ideal fluid limit. The form of $\kappa_{\rm lmfp}$ can be argued on dimensional analysis, up to a constant \citep{Lynden-Bell1980}. SIDM applications use the parameterization $\kappa_{\rm lmfp} = 0.27\beta nv^{3}_{\mathrm{rms}} \sigma_m k_{B}/(G)$ (depending on number density $n$, velocity $v$, the dark matter particle mass $m$ via $\sigma_m\equiv\sigma/m$, and the Boltzmann and Newton constants, $k_B$ and $G$), with the scaling constant $\beta$ determined by comparing the analytic results to SIDM N-body simulations \citep{Essig_2019, Yang:2022zkd,Outmezguine:2023}. Notice that the ratio of conductivities follows the estimated Knudsen number:
\begin{equation}
    \frac{\kappa_{\rm smfp}}{\kappa_{\rm lmfp}}\propto\frac{{\rm Kn^2_{gravothermal}}}{\beta}\,.
\end{equation}

With the ansatz in Eq.(\ref{eq:kappa}) as well as a prescription for choosing $\beta$ and the assumption of spherical symmetry, the system of differential equations for the evolution of a halo's radial profile $\rho(r)$, total halo mass $M_h$, luminosity $L$, and the 1D root-mean square velocity averaged over the Maxwell-Boltzmann distribution ($v_{\mathrm{rms}}(r)$) can be solved (see \cite{Yang2023} for precise details). 

To compare the collapse of different halos, with different self-interaction cross-sections, the solutions can be conveniently written in terms of a set of dimensionless, scale-free variables. For example, the radial density profile can be written as $\hat{\rho}(r, t)=\frac{\rho(r,t)}{\rho_s}$ with the constant $\rho_s$ from the Nevarro-Frenk-White (NFW) profile of the initial conditions,
\begin{equation}
    \rho(r, t_0)=\frac{\rho_s}{\frac{r}{r_s}\left(1+\frac{r}{r_s}\right)^2}\,.
    \label{eq:nfw_profile}
\end{equation}
The convenient dimensionless clock is $\hat{\sigma}_m\hat{t}\equiv\left(\frac{\sigma}{m}\rho_sr_s\right)\frac{t}{\sqrt{4\pi G\rho_s}}$. 

\begin{figure}
	\includegraphics[width=\columnwidth]{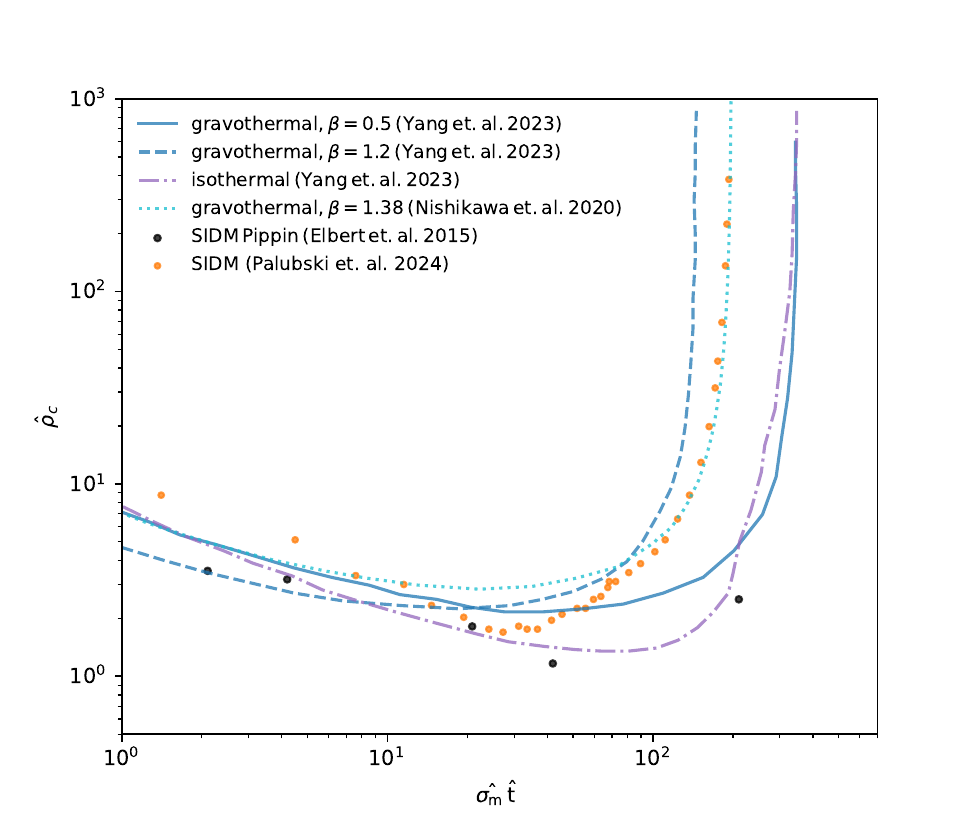}
    \caption{The collapse curves for self-interacting dark matter halos, evolved using various methods and normalized following the methodology in \protect\cite{Yang2023}. The solid blue and dashed blue lines show results from the gravothermal semi-analytic method (with $\beta = 0.5$ and $\beta = 1.2$, respectively), and the dot-dashed purple curve shows an iso-thermal result, all from \protect\cite{Yang2023}. The dotted cyan curve shows a gravothermal model from the work of \protect\cite{Nishikawa_2020}, with $\beta = 1.38$ and interaction cross-section per unit mass of $50\, \mathrm{cm}^{2}/\mathrm{g}$ . The solid circles show numerical results from SIDM simulations, including the Pippin simulation in black (from \protect\cite{Elbert2014}). Orange circles show a $10^6$ particle halo collapsing at redshift $z=31$, with a spline scattering implementation, and $\kappa=0.002$ (from \protect\cite{Palubski:2024ibb}).\label{fig:reviewSIDMcurves}}
\end{figure}

Figure \ref{fig:reviewSIDMcurves} reproduces several examples of gravothermal collapse results from the literature. Each curve shows the evolution of the scale-free central density, $\hat{\rho}_c = \rho\big|_{r=0} / \rho_{s}$, with rescaled time $\hat{\sigma}_m\hat{t}$. Results for this gravothermal approach for three different values of $\beta$ are shown in blue (from \cite{Yang:2022zkd}) and cyan (from \cite{Nishikawa_2020}). 

The purple dot-dashed curve in Figure \ref{fig:reviewSIDMcurves} shows an example result from \cite{Yang:2022zkd}, using an alternative semi-analytic approach, the isothermal method \citep{Yang2023}. In this technique, the matching is done spatially, assuming that the outer part of the halo is described by an NFW profile, while the inner part is a direct solution of the Boltzmann equation. 

The final two examples shown in Figure \ref{fig:reviewSIDMcurves} are numerical simulations of SIDM, where stochastic self-scattering of N-body particles mimics the effect of DM self-interactions. This is the Ansatz of standard SIDM numerical techniques, which assumes that the scattering dynamics of the microscopic particles can be applied on the macroscopic scale. For example, in the treatment set out by \cite{Rocha2013, Robles_2017}, a scattering rate of an N-body particle $j$, with mass $m_{p}$, off of a nearby target particle $i$ is given by
\begin{equation}
\label{eq:SIDM_scatter}
\Gamma(i|j) = (\sigma_{\rm elastic} / m )\ m_{p} |\mathbf{v}_{i} - \mathbf{v}_{j} | g_{ij}\,,  
\end{equation}
where $\mathbf{v}_{i}$, $\mathbf{v}_{j}$ are the particle velocities and
\begin{equation}
\label{eq:kernel}
g_{ij} = \int^{max(h_{i}, h_{j})}_{0} d^{3}\mathbf{x}'W(|\mathbf{x'}|, h_{i})W(|\delta \mathbf{x}_{ij} + \mathbf{x}'|, h_{j} )
\end{equation}
accounts for the mass density overlap of the two particles' cells, with $h_{i}$ the self-interaction smoothing length. 

The probability that two particles scatter during a time step $\delta t$ and the total probability of an interaction occurring between the two particles are then (respectively), 
\begin{equation}
P(i|j) = \Gamma(i|j) \delta t, \quad P_{ij} = \frac{P(i|j) + P(j|i)}{2}\,.    
\end{equation}
In the case that particles interact, an elastic scattering `kick' is applied, changing the particle velocities as
\begin{equation}
\begin{aligned} 
\label{eq:SIDMkick}
\mathbf{v}'_{i} &= \mathbf{v}_{CM} + \frac{m_{j}}{m_{i} + m_{j}} V \mathbf{e},\\
\mathbf{v}'_{j} &= \mathbf{v}_{CM} - \frac{m_{i}}{m_{i} + m_{j}} V \mathbf{e}\,.  
\end{aligned}
\end{equation}
Here $\mathbf{v}_{\mathrm{CM}}$ is the center of mass velocity of the colliding particles, $V$ is the relative speed between the two particles, and $\mathbf{e}$ is a randomly chosen direction. Angular momentum is not conserved in this SIDM treatment, while total energy and linear momentum are.

In practice, there are several simulation parameters that must be adjusted both for the scattering algorithm and the usual N-body dynamics. Different choices can lead to significantly different results. The effects of varying the mass resolution, force-softening length, and time-stepping criteria were studied by \cite{Mace:2024uze} for the SIDM implementation in the Arepo code, and for three different SIDM algorithms using GIZMO and Arepo in \cite{Palubski:2024ibb}, following an earlier comparison by \cite{Meskhidze:2022}. 

The significant difference in collapse times shown in Figure \ref{fig:reviewSIDMcurves}, and more generally in the detailed results of \cite{Meskhidze:2022, Mace:2024uze, Palubski:2024ibb}, demonstrates that there is not yet a clear consensus on the precise collapse behavior. In addition, existing numerical SIDM algorithms become very slow in high-density regimes, as the timestep must drastically shrink in order for the scattering calculation to remain in its regime of validity. The semi-analytic techniques calibrated on simulations are very efficient for symmetric halos, but are not flexible enough to handle variations in halo morphology or angular momentum. Both of these short-comings can be addressed by a technique that can connect the N-body regime to the hydrodynamic description. This is what we develop next.

\subsection{An SIDM-hydro hybrid approach}
The methods outlined above offer phenomenological insight into the evolution of self-interacting halos. When the evolution of the halo profile is examined, both semi-analytic and numerical treatments show the initial broadening and flattening of the central density, after which the continuous increase of density and thermal energy at the center of the structure leads to gravothermal collapse. What the various methods do not converge on is the precise time-scale over which the collapse occurs (as was demonstrated in Figure \ref{fig:reviewSIDMcurves}) or the physical state (is it a black hole?) of the halo's dense center as its density spikes. This motivates continued refinements of numerical techniques, in particular for the high-density limit. We next re-examine some of the conventions leading into the derivation of the fluid equations in the frequent scattering limit, laying the groundwork for our proposed method.

The challenges we are facing is as follows. First, outside of the small-Knudsen number limit, integrating the Boltzmann equation with elastic scattering is numerically difficult. The gravothermal approach relies on limiting assumptions such as hydrostatic equilibrium, sustained spherical symmetry, and a form for the thermal conductivity. Meanwhile, maintaining adequate spatial resolution and accurate energy conservation is a challenge for N-body methods at intermediate Knudsen numbers.

The challenge of transitioning from the small to frequent-scattering limit is not unique to SIDM, and other authors \citep{Klar:1995, Tidriri:2001} have proposed boundary matching across domain decomposition (akin to the isothermal semi-analytic model's implementation) or a parameterization between the two solutions using an artificial connecting function across a buffer zone \citep{DegondJin:2005}. Others have suggested solving for both solutions across the entire domain and committing to one or the other according to a well-defined local criteria \citep{Tiwari1998}.  

Our approach seeks to combine aspects of previously proposed kinetic-to-fluid equation coupling methods to imbue existing SIDM techniques with additional fluid physics in the small-Knudsen limit. It is not a full, first-principles solution and requires some additional data to determine the best choice of parameters. A previous suggestion to incorporate a fluid description was made in \cite{Kummer:2019yrb}, who used N-body particles (a pressure-less fluid treatment) for the dark matter, but with a prescription for introducing effective interactions that relied on solving a diffusion equation for pressure-less fluids. Their technique makes use of the thermal conduction SPH module from \texttt{GADGET-2} to simulate an isothermal core in isolated dwarf galaxy halos, though not the full gravothermal re-collapse. The method of \cite{Kummer:2019yrb} also does not incorporate the scattering mechanisms of traditional SIDM methods, though is intended to do so present an comprehensive picture.

To motivate our hybrid method, consider the Boltzmann equation,
\begin{equation}
\label{eq:Boltzmann}
 \frac{\mathrm{d}f}{\mathrm{d}t}= \frac{\partial f}{\partial t} + \mathbf{v}\cdot \nabla f - \nabla \Phi \cdot \frac{\partial f}{\partial v} = \mathbb{C}[f]\,,
\end{equation}
where $f(\mathbf{x}, \mathbf{v}, t)$ is the phase-space distribution function, $\mathbf{x}$ and $\mathbf{v}$ are, respectively, position and velocity of particles, $\Phi$ is the gravitational (assuming the only inter-particle forces are gravitational), and $\mathbb{C}[f]$ the collisional term.

The moments of the Boltzmann equation, in terms of collisionally invariant quantities $\mathbf{Q}(\mathbf{v})$, are often more practical to work with. The first moment for $\mathbf{Q}(\mathbf{v}) = m\mathbf{v}$ yields the familiar momentum equation,

\begin{equation}
\label{eq:FirstMoment}
\frac{\mathrm{d} u_j}{\mathrm{d} t} = \frac{\partial u_j}{\partial t}+u_i\frac{\partial u_j}{\partial x_i}=\frac{1}{\rho} \frac{\partial \sigma_{ij}}{\partial x_i} \, -  \frac{\partial \Phi}{\partial x_j}.
\end{equation}
Here $u_{j}$ is the bulk, or streaming, velocity  of the fluid and $\sigma_{ij} = -\rho \langle w_{i}w_{j} \rangle_{f}$ is the stress tensor, where $\rho$  is the density, $w_{i} \equiv v_{i} - u_{i}$ is the random local motion and $\langle \rangle_{f}$ the phase-space velocity average quantity over the phase-space distribution $f$. In the limit of frequent collisions, the stress-tensor can be instead written in terms of the pressure and viscosity $\sigma_{ij} = -P \delta_{ij} + \tau_{ij} $. Providing an additional equation of state $P(\epsilon, \rho)$ allows the moment equations to form a closed set, where $\epsilon$ is the internal energy, which is the subject of the second moment equation:
\begin{equation}
\label{eq:SecMoment}
\frac{\mathrm{d}\epsilon}{\mathrm{d}t} = \frac{\partial \epsilon}{\partial t} + u_{i} \frac{\partial \epsilon}{\partial x_{i}} = -\frac{P}{\rho} \frac{\partial u_{i}}{\partial x_{i}}.
\end{equation}
The above drops thermal flux and anisotropic stress terms to arrive at the usual fluid approximation implemented in SPH simulations.

In the first equality in Eq.(\ref{eq:FirstMoment}) and (\ref{eq:SecMoment}), we included both the Lagrangian $\frac{\mathrm{d}}{\mathrm{d}t}$ and Eulerian $\frac{\partial}{\partial t} + \mathbf{v}\cdot \nabla$ specification of the fluid flow field. The choice of reference frame has notable implications for the numerical prescription of the equations with each having distinct advantages (for reviews of the various methods see, for example, \cite{Trac:2003,Agertz:2007,Springel:2010, Hu:2023}). Our proposed hybrid method presupposes a Lagrangian formulation, both to make it compatible with existing Lagrangian SIDM models, and to avoid the problems that arise with the implementation of momentum conservation across fluid cell boundaries indelible to Eulerian Godunov schemes. We return to the associated shortcomings of Lagrangian Smoothed Particle Hydrodynamics (SPH) in later discussion.

SIDM methods, such as in \cite{Rocha2013}, derive their Lagrangian solution to the Boltzmann equation by assuming that the macroscopic simulation particles, representing an average over the distribution function, can be well described by the same equation as the individual fluid particles. If the macroscopic discretization of $f$ is represented by $\hat{f} = \sum_{i}(M_{i}/m)W(|\mathbf{x} - \mathbf{x}_{i}; h_{i}|)\delta^{3}(\mathbf{v} - \mathbf{v}_{i})$, then the motion equation of an individual simulation particle with mass $M_{i}$ can be written as 
\begin{equation}
\frac{\mathrm{d}u_{i}}{\mathrm{d}t} = \mathrm{Kick}_{\mathrm{SIDM}, i} - \frac{\partial \Phi}{\partial x_{i}}.
\end{equation}
The scattering operator $\mathrm{Kick}_{\mathrm{SIDM},i}$ is a velocity kick, as described in Eq.(\ref{eq:SIDM_scatter}) - Eq.(\ref{eq:SIDMkick}). 

\begin{figure*}
	\includegraphics[width=\textwidth]{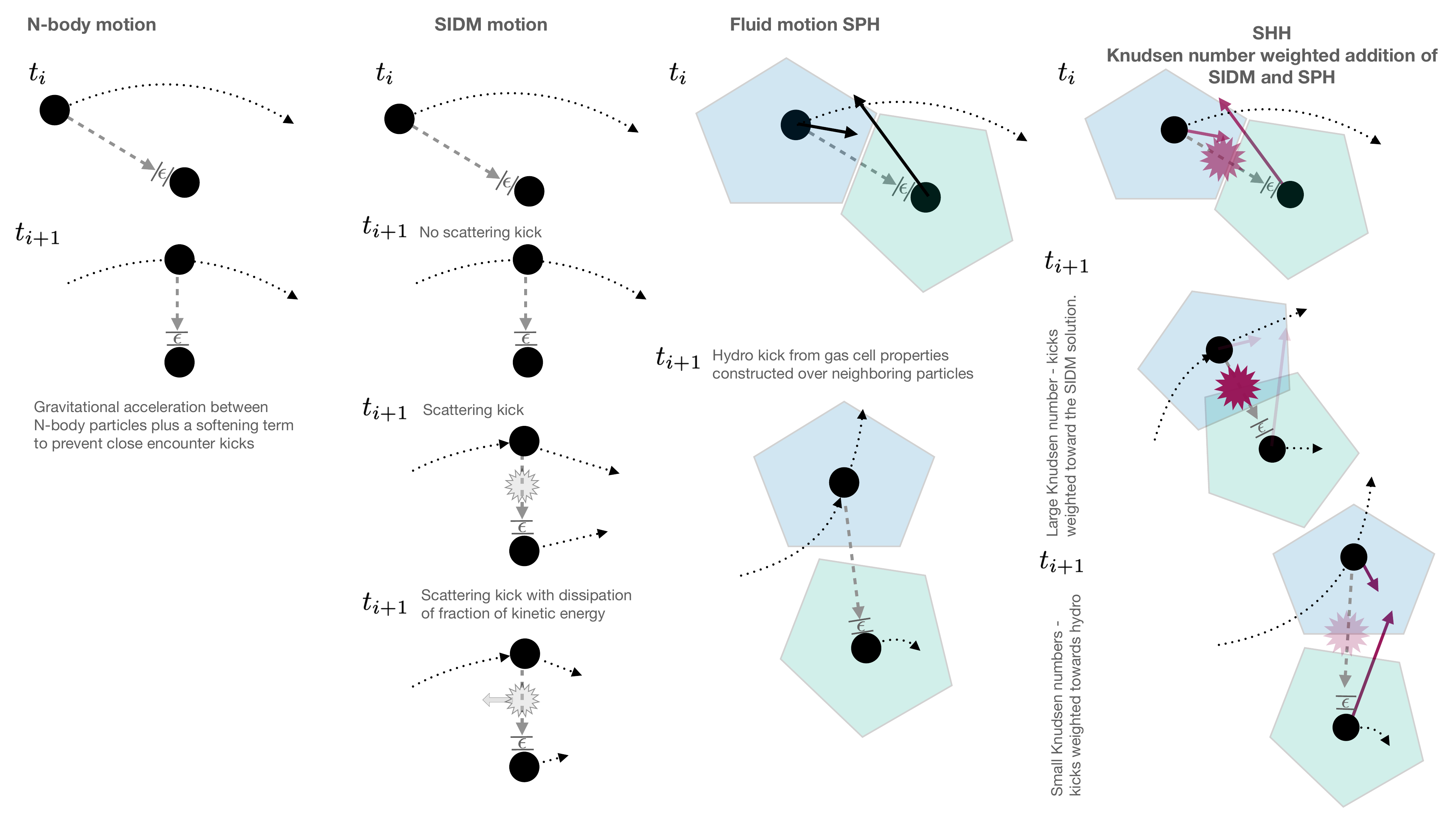}
    \caption{Overview of dynamic methods relevant to simulations. Left to right: N-body only, numerical SIDM, SPH, and our newly proposed hybrid method. In the gravity-only N-body scheme, acceleration due to gravitational forces (gray, dashed arrows) is calculated (including the force softening parameter $\epsilon$) and the trajectories (dotted black arrows) are updated via the subsequent momentum kick over a discreet timestep $t_{i+1}$. In SIDM (second column), a local interaction probability governs whether a scattering kick is applied (star shape). The kick conserves energy and momentum but applies a random choice in the direction. In the dissipative SIDM case (bottom diagram) some of the total kinetic energy of the collision is lost. In the fluid SPH method (third column), a nearest neighbor particle cell (green and blue polygons) is constructed from which local hydro properties are calculated. There are used to update the local hydro forces (black, solid arrows) and these are used to update the momentum kicks in additional to those due to gravitational forces. Our newly proposed hybrid model (fourth column) implements a weighted product (presented via the opacity of the purple kick components) of the SPH and SIDM method, with large Knudsen numbers weighted towards the SIDM solution (middle diagram) and small Knudsen numbers towards SPH hydro (bottom diagram). 
    \label{fig:SIDM_graphic}}
\end{figure*}

We note that functionally this solution is very alike the first moment equation, Eq.(\ref{eq:FirstMoment}), with the pressure term $\partial P_{ij}/\partial x_{i}$ replaced by the SIDM kick $\mathrm{Kick}_{\mathrm{SIDM}} = \mathbb{C}[f]$. However, by construction, the SIDM Ansatz will fail when the mean free path of the physical particles $\lambda_{\rm mfp}$ becomes appreciably smaller than the size of the coarse-grained particle $l_c$. Instead, the internal energy defined by the random motion of the physical particles which comprise each simulation particle needs to be taken into account. If the distribution function is close to equilibrium, this internal energy can be described by a single number: the temperature of the Maxwell-Boltzmann distribution. We thus propose augmenting the SIDM solution with the full fluid description in that limit. We do not attempt a spatial domain decomposition of the solution to the Boltzmann equation due to the dynamic, inhomogeneous environment of a halo. Instead, we consider a combination of both solutions across the halo's domain, weighted by a bridging function of the local Knudsen number (illustrated in Fig.\ref{fig:SIDM_graphic}). Then the first moment equation, in the Lagrangian hybrid technique, is
\begin{equation}
\label{eq:accel_with_bridging}
\frac{\mathrm{d} u_{i}}{\mathrm{d}t} =  h(\rho) \mathrm{Kick}_{\mathrm{hydro},i} + (1 - h(\rho)) \mathrm{Kick}_{\mathrm{SIDM}, i} - \frac{\partial \Phi}{\partial x_{i}}\,,
\end{equation}
where $h(\rho): \mathbb{R} \rightarrow \mathbb{R}$ is the continuous, monotonic bridging function with limits $h(\rho)|_{\mathrm{Kn} \rightarrow 0} = 1$ and $h(\rho)|_{\mathrm{Kn} \rightarrow \infty} = 0$. The hydro kick, $\mathrm{Kick}_{\mathrm{hydro},i} = -\frac{1}{\rho}\frac{\partial P}{\partial x_{i}}$ and SIDM kick, are the locally rescaled terms of non-modified dynamic equations. We have formed Eq.~\ref{eq:accel_with_bridging} such that if the two kick terms are equal the impact of $h$ vanishes. The formulation also resembles a ``modified pressure,'' but because we do not include a $\partial h/\partial x_i$ term this interpretation cannot be taken completely literally. (Given that the fluid cells are numerically constructed to assign constant hydrodynamic quantities to their associated particles, it is reasonable to assume $h(\rho)$ only varies to a negligible degree across the cell.)

The modified energy equation is a little more subtle, as we are looking for an energy term that corresponds directly to the hydro force represented by the modified change in momentum, to ensure overall energy conservation. As a first order pass, we will here look for a simple interpretation with an eye to minimizing the required modification of existing numerical SPH implementations. We adopt
\begin{equation}
\frac{\mathrm{d}\epsilon}{\mathrm{d}t} =  -h(\rho )\frac{P}{\rho}\frac{\partial u_{i}}{\partial x_{i}},
\end{equation}
which makes apparent that the ``modified pressure'' interpretation is valid if the spatial derivatives of $h$ are negligible.
Since the N-body SIDM algorithm makes no provisions for change in internal energy during scattering events, there is no contribution from this component in our modified method. (Technically speaking the contribution to the internal energy change from the SIDM scattering term is $(1 - h(\rho)) \cdot 0$, as the technique makes no provision for the internal energy of the fluid cell and assumes all energy change is purely facilitated via the bulk velocity term. Realistically we should expect some modification to $w_{i}$ even in the low and intermediate scattering regime, and this represents an inherent shortcoming of traditional SIDM scattering methods.)

At this point we have not discussed the more general, frequent-collision solution to the Boltzmann equation. The Navier-Stokes equation includes non-ideal fluid terms which can be derived as an expansion around equilibrium Maxwell-Boltzmann solution of the phase-space distribution function $f$, using (nearly) the Knudsen number as the expansion parameter \citep{ChapmanCowlingBook,enskog1917}:
\begin{equation}
\label{eq:NavierStokes}
\begin{split}
\frac{\mathrm{d}u_{i}}{\mathrm{d}t} &= -\frac{1}{\rho}\frac{\partial P}{\partial x_{i}} +  \frac{1}{\rho}\frac{\partial}{\partial x_{j}} \Bigl[ \mu \left( \frac{\partial u_{i}}{\partial x_{j}} + \frac{\partial u_{j}}{\partial x_{i}} - \frac{2}{3} \delta_{ij}\frac{\partial u_{k}}{\partial x_{k}} \right)\Bigr] \\ 
&+ \frac{1}{\rho} \frac{\partial}{\partial x_{i}} \left( \eta \frac{\partial u_{k}}{\partial x_{k}}\right) -  \frac{\partial \Phi}{\partial x_{i}},
\end{split}
\end{equation}

\begin{equation}
\label{eq:NonIdealEnergy}
\frac{\mathrm{d}\epsilon}{\mathrm{d}t} = - \frac{P}{\rho}\frac{\partial u_{k}}{\partial x_{k}} + \frac{1}{\rho} \tau_{ik} \frac{\partial u_{i}}{\partial_{k}} - \frac{1}{\rho} \frac{\partial F_{\mathrm{cond}, k}}{\partial x_{k}},
\end{equation}
where $\mu$ and $\eta$ are the coefficients of shear and bulk viscosity respectively, $\Upsilon = \tau_{ik}\frac{\partial u_{i}}{\partial x_{k}}$ is the rate of viscous dissipation. $F_{\mathrm{cond}, k} = \hat{\kappa} \frac{\partial T}{\partial x_{k}}$ is the conductivity, with $\hat{\kappa}$ denoting the thermal conduction rate (we use $\hat{\kappa}$ to distinguish it from the SIDM tolerance parameter $\kappa$ us throughout this work).

Although we do not include viscosity and thermal conductivity terms in the remained of this work, these terms would form a natural extension to the formalism described above. 

\subsubsection{The bridging function}
\label{sec:bridging}
Having established the conceptual foundation for the bridging function above, we now provide a prescription for its implementation in simulations. Let us first define the Knusden number using the simulation variables and in reference to the fluid cell scale.  We approximate the numerical cell volume to be spherical, with total mass enclosed defined by simulation particles in the cell, set by the nearest neighbor number $N_{\mathrm{ngbr}}$, and simulation particle mass, $N_{\mathrm{ngbr}} \cdot M_{\mathrm{sim}}$\footnote{We use this form as it displays the parametric dependence on simulation parameters. We verified that the cell radius calculated in this manner scales with the kernel smoothing length up to a constant. However, one could instead use the volume, including the kernel, already calculated in GIZMO.}, so that 
\begin{align}
\label{equ:Knudsen_sims}
  {\rm Kn}_{\mathrm{sim}}(\rho) =  \frac{\lambda_{\mathrm{mfp}}}{l_{\mathrm{c}}} &=\frac{1}{\sigma n}\left(\frac{N_{\rm ngbr}M_{\rm sim}}{\rho \frac{4}{3}\pi}\right)^{-1/3}\\\nonumber
    &=\frac{1}{\sigma_m\left(\frac{N_{\rm DM}m}{\rho}\right)}\left(\frac{N_{\rm ngbr}M_{\rm sim}}{\rho \frac{4}{3}\pi}\right)^{-1/3}\\\nonumber
   &=\frac{1}{\sigma_m\rho}\left(\frac{4\pi\rho}{3N_{\rm ngbr}M_{\rm sim}}\right)^{1/3}\\\nonumber
   &= \frac{1}{\sigma_{\mathrm{m}}\rho^{2/3}} \left(\frac{4 \pi}{ 3 N_{\mathrm{ngbr}} M_{\mathrm{sim}}} \right)^{1/3}\,,
\end{align}
where $\sigma_{\mathrm{m}}$ is the dark matter self-scattering cross-section per unit DM particle mass, and $\rho$ is the mass density of the DM SPH particles, which is equivalent to the physical mass density of microphysical dark matter particles. This Knudsen number is dependent both on the physical properties of the SIDM particle via $\sigma_{\mathrm{m}}$, and on the configuration of the simulation via $N_{\mathrm{ngbr}}$ and $M_{\mathrm{sim}}$. A dependence on simulation parameters is typical of hybrid schemes similar to ours. One could extend the algorithm to use a minimum of the numerical Knudsen number and a Knudsen number defined by a purely physical scale \citep{Liu:2019}. While the choice used above is natural, it would be worthwhile to explore the implications of more closely aligning ${\rm Kn}_{\mathrm{sim}}$ and ${\rm Kn}_{\mathrm{gravothermal}}$. We leave this for future work.

Clearly, ${\rm Kn}_{\mathrm{sim}}$ is also parametrically distinct from the Knudsen number defined in the gravothermal collapse literature, Eq.(\ref{eq:KnGravothermal}). The emphasis on the the numerical Knudsen number in this work is motivated as follows. No method can capture flow features which vary on scales small compared to the simulation resolution, and both kinetic and fluid methods are on equal footing when flow features hardly vary at the resolution scale. The numerical Knudsen number, comparing the mean-free-path to the scale of the fluid cell, gives us a proxy for how far the fluid cell is from local collisional equilibrium and we use this to weight contributions from the two methods. On the other hand, if the unresolved features are in fact near the continuum regime (as, for example, in collisional shocks) a fluid description will lead to an accurate ``coarse grained'' solution at the resolution scale. We expect this near-continuum assumption to hold approximately (though not exactly, see \cite{gurian:2025}) in the gravothermal collapse problem. Further, in this resolution-limited regime, the fluid description will be less noisy than the N-body SIDM prescription. Finally, in completely unresolved regions it is reasonable to fall back to the fluid solver due to the greatly reduced computational cost.

\begin{figure}
	\includegraphics[width=\columnwidth]{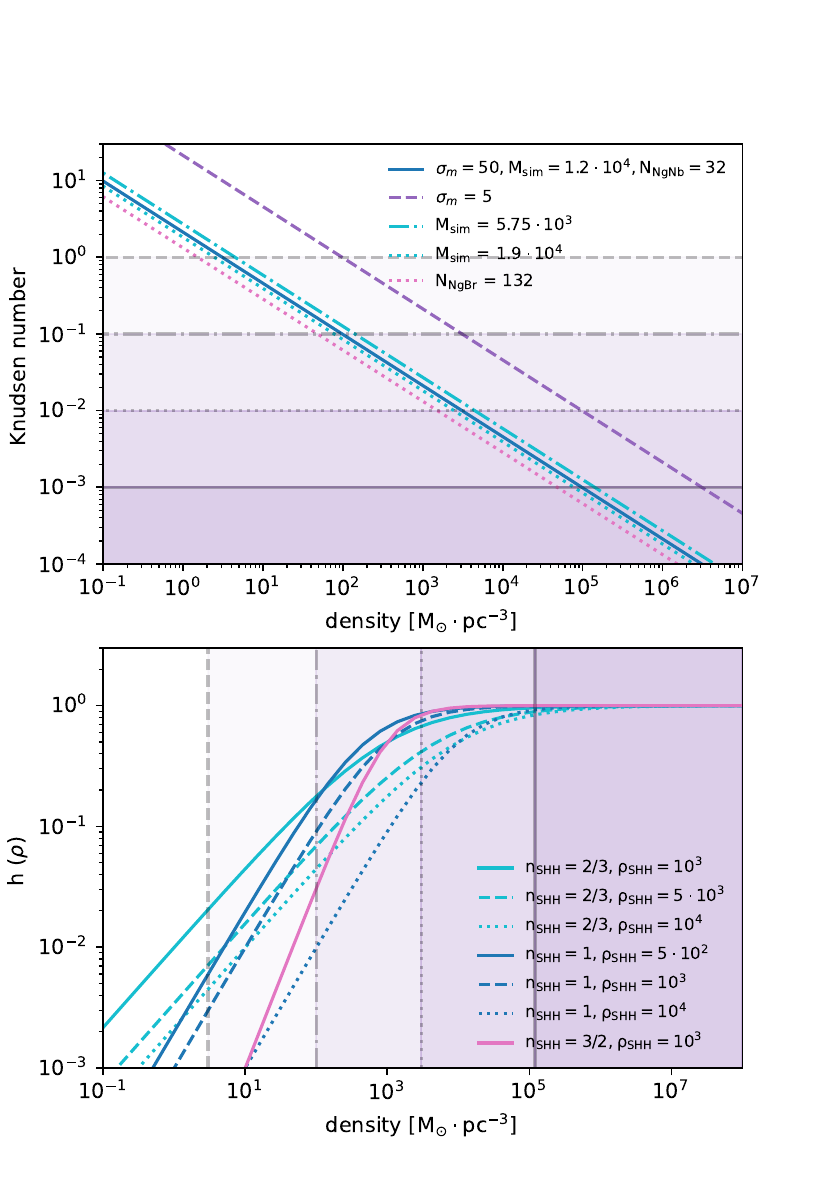}
    \caption{{\bf Top panel:} The simulation Knudsen number, Eq.(\ref{equ:Knudsen_sims}), for various dark matter particle and simulation parameters. In each scenario the DM hydrodynamic gas cell is constructed via an SPH kernel for a fixed nearest neighbor number and the cell size derived therefrom. The blue curves show the Knudsen number for a SIDM cross-section of $50$ [cm$^{2}$/ g], with an assumed DM particle mass of $10$ M$_{\mathrm{proton}}$, simulation mass of $1.2 \cdot 10^4$ $\mathrm{M}_{\odot}$, and nearest neighbor number ($\mathrm{N}_{\mathrm{ngbr}}$) of 32. The remaining curves use the same parameters, except for a single change as listed in the legend. The dashed purple curve has SIDM cross-section of $5$ [cm$^{2}$/ g], the cyan curves have simulation mass, $\mathrm{M}_{\mathrm{sim}},$ of $5.75 \cdot 10^3$ $\mathrm{M}_{\odot}$ and $1.9 \cdot 10^4$  $\mathrm{M}_{\odot}$ for the dot-dashed and dotted lines respectively. (These correspond to the simulation masses of our low and high resolution halos.) Lastly the pink, dotted line shows the Knudsen number for cells constructed from 132 nearest neighbor particles. {\bf Bottom panel:} The effect of changing the power-law scaling, $\mathrm{n}_{\mathrm{SHH}}$, and the threshold density, $\rho_{\mathrm{SHH}}$ , in the bridging function, Eq.(\ref{eq:fKn}). Knudsen numbers corresponds to values for our high resolution halo, or the cyan dot-dashed curves in the plot above. \label{fig:Knudsen_functions}}
\end{figure}

The top panel of Figure \ref{fig:Knudsen_functions} shows ${\rm Kn}_{\mathrm{sim}}(\rho)$ for several choices of parameters. The solid blue curve uses baseline parameters, and all others show the effect of varying a single parameter to the value indicated in the legend. The buffer region in which the solution of the Boltzmann equation transitions from kinetic to hydro-dynamic extends from ${\rm Kn}\sim 1$ down to ${\rm Kn}<10^{-4}$. The purple shaded bands in each panel indicate the domain over which the bridging function goes from nearly 0 (lightest color) to nearly 1 (darkest color). Following the literature, we take regions for which $\mathrm{Kn} \lesssim 10^{-3}$ to be fully in the hydro regime.

We choose a two-parameter form of the bridging function,
\begin{equation}
h(\rho) = \frac{1}{1 +  \left( \frac{\rho_{\mathrm{SHH}}}{\rho}  \right)^{n_{\mathrm{SHH}}}}\,.
\label{eq:fKn}
\end{equation}
The parameter $n_{\mathrm{SHH}}$ determines how rapidly the contribution of the hydro term falls off with decreasing density. We will explore a range of values $2/3\leq n_{\mathrm{SHH}} \leq 3/2$ below. The parameter $\rho_{\rm SHH}$ specifies the threshold density where $h(\rho)$ starts to deviate from 1, down-weighting the hydrodynamical pressure and adding an SIDM N-body kick component to the change in momentum. This threshold should be chosen such that the value of the bridging function is very nearly one at the density, $\rho_*$, where the Knudsen number is $10^{-3}$. The threshold simulation density above which the hydrodynamical description should apply is obtained by solving for $\rho$ in Eq.(\ref{equ:Knudsen_sims}) with ${\rm Kn_{sim}}=10^{-3}$:
\begin{equation}
\label{eq:ComputerhoSHH}
    \rho_*=\left(\frac{10^3}{\sigma_{\mathrm{m}}}\right)^{3/2} \left(\frac{4 \pi}{ 3 N_{\mathrm{ngbr}} M_{\mathrm{sim}}} \right)^{1/2}\,.
\end{equation}
Notice that as we change the resolution, holding everything else constant, $\rho_{\rm SHH}$ must change. This is because ${\rm Kn}_{\rm sim}$ compares the physical mean free path to the simulation particle resolution, in order to determine whether the scattering of physical particles would be sufficient to assign an internal energy and other hydrodynamical quantities to the simulation particles. In other words, if one instead adaptively increased resolution, in principle the kinetic dynamics would remain accurate for a longer time.

Let us explore varying $\rho_{\rm SHH}$ by slightly relaxing how closely $h(\rho|_{\mathrm{Kn_{sim}}=10^{-3}}\equiv\rho_*) =1$ is enforced. If we require $h(\rho_*)=1-\delta$, the same threshold density for the turn-on of the hydro terms can be implemented consistently across different choices in the steepness of $n_{\mathrm{SHH}}$ by requiring the parameter $\rho_{\rm SHH}$ to satisfy
\begin{equation}
    \rho_{\rm SHH}=\rho_*\left(\frac{\delta}{1-\delta}\right)^{\frac{1}{n_{\rm SHH}}}\,.
\end{equation}

The bottom panel of Figure \ref{fig:Knudsen_functions} shows the bridging function for various choices of $\rho_{\mathrm{SHH}}$ and $n_{\mathrm{SHH}}$. The purple bands show the different magnitudes of $\mathrm{Kn}$ computed from the solid blue curve in the top panel. The cyan, blue, and pink families of curves correspond to bridging functions with $n_{\mathrm{SHH}} = 2/3, 1,$ and $3/2$ respectively. Varying the parameters changes the contribution of the hydro term at lower densities, which in turn affects the halo dynamics. After discussing our implementation of the hybrid method in GIZMO in the next section, we will present the different outcomes for these bridging functions in Section \ref{sec:sims}. 

\section{SHH implementation in simulation}
\label{sec:implementation}
We implement the SHH scheme within GIZMO \citep{Hopkins2015}. Since our modifications presuppose a Lagrangian formulation, we make use of GIZMO's native pressure-SPH module \citep{Hopkins_2012}\footnote{GIZMO's novel meshless hydrodynamic solver also retains a similar functional form as the Lagrangian equation. However, it also makes use of a Riemann solver, which would require more considered modification to implement our scheme correctly. The Riemann solvers in question do exhibit advantages over SPH schemes, especially in the context of discontinuities and shocks in the fluid, so this issue may be worth revisiting in future work.}. All our simulations use GIZMO compiled with a cubic spline kernel function, polytropic index $\gamma = 5/3$ in the ideal gas law, and conservative ($0$) slope limiter tolerance. 

We use adaptive gravitational softening for all particles since we attach the SIDM loop to gas particles instead of the traditional N-body types. This is recommended practice for gas particles so that the Poisson equation is solved on the same particle distribution as the Euler equation, to avoid numerical artifacts and non-converging solutions. The gravitational, hydro, and SIDM processes for all our simulations therefore share the same kernel softening length. 

The tolerance parameter $\eta$ (a parameter of interest in \cite{Palubski:2024ibb}) is set to $0.001$ for all halos in this work, while $\kappa$, the SIDM time stepping criterion, is allowed to vary between simulations. The SIDM portion of the dynamic calculation comes from the GIZMO native module by \citep{Rocha2013}. 

\subsection{Modifications to GIZMO}
\label{sec:GIZMO_mods}
The additional modifications made to the SIDM module, and other loops of GIZMO, to accommodate the hybrid method are detailed in this section and outlined in Figure \ref{fig:SIDM_graphic}. The key idea is to attach SIDM attributes to a standard gas simulation particle. The kicks are modified by the bridging function, Eq.(\ref{eq:fKn}), according to the local density, which is taken to be a proxy for the Knudsen number via Eq.(\ref{equ:Knudsen_sims}). 
Equivalent adjustments are made to the corresponding time-step criteria. 

\subsubsection{Dynamic, energy, and SIDM scattering equation modifications}
\label{sec:modified_equations}
Simulation particles are dynamically evolved via momentum kicks, calculated for each time step from the configuration of the ensemble of the surrounding particles. Following Eq.(\ref{eq:accel_with_bridging}), we note that the gravitational acceleration kick ($\texttt{kick[i,j]}_{\texttt{g}} \texttt{= mass\textunderscore pred}\, * \,\texttt{P[i].GravAccel[j]}\, * \,\texttt{dt\textunderscore gravkick} $) remains unchanged. 

The hydro kick is suppressed via our bridging function ($\texttt{kick[i,j]}_{\texttt{mod hydro}}\texttt{ = mass\textunderscore pred}\, * \, \texttt{h(SphP[i].Density)}    \, * \,\texttt{SphP[i].HydroAccel[j]}\, * \,\texttt{dt\textunderscore hydrokick}$). This is done directly in the $\texttt{kicks.c}$ file, after the hydro terms have been calculated in the standard manner in the hydro loop. Here $\texttt{mass\textunderscore pred}$ should be taken as the fixed simulation particle mass. The notation arises from GIZMO allowing for scenarios in which the masses are not fixed such as the meshless finite volume method. The index $\texttt{i}$ is the simulation particle the kick is acting upon, while $\texttt{j}$ denotes the index of the Cartesian coordinate of momentum.

The modified hydro component then reads,
\begin{equation}
\texttt{dp[j] += kick[P[i,j]]} _{\texttt{grav}} \texttt{+ kick[SphP[i,j]]}_{\texttt{mod hydro}}.
\end{equation}

In order to attach the SIDM loop to our SPH particles, we compile the code with the SIDM flag to also act on gas particles. An additional small change is needed here to comment out the density calculation native to the SIDM module. We simply replace it with the density already associated with the gas particle. The kernels for the hydro and SIDM kick are thus the same for our simulations. The SIDM kick is added after the gravitational and hydro kicks, and only if the scattering probability condition is met. We modify the SIDM component by applying $(1 - h(\rho))$ to that scattering condition, so that 
\begin{equation}
\texttt{gsl\textunderscore rng\textunderscore uniform < (1 - h(SphP[i].Density))}\, * \, \texttt{prob}.
\end{equation}
The choice of implementation is here perhaps less obvious than in the hydro term. In the development process we also trialed applying the $h(\rho)$ suppression term directly to the velocity kick to match its hydro counterpart. However we found several undesirable side-effects to this approach, namely less robust energy conservation and a time bin crash due to the scattering condition still being met despite there being no change in the actual particle dynamics.

There is commensurate modification of the internal energy terms to account for the reduced hydro force.  $\texttt{DtIntEnergy = SphP[i].DtInternalEnergy} \, * \, \texttt{dt\textunderscore hydrokick}$

\begin{equation}
\texttt{du\textunderscore tot = h(SphP[i].Density))}\, * \, \texttt{DtIntEnergy + dEntGravity}    
\end{equation}
The SIDM loop has no internal energy considerations and needs no further modification.

\subsubsection{Time stepping criteria}
\label{sec:timestepping}
As is the case for all numerical methods, the time step in traditional SIDM method is informed by balance between expedience and the faithfulness of the solution. There are two primary factors driving the time-step size in the standard method: self-gravity and the scattering criterion $\kappa$. The former is set by the acceleration of the particle and the force softening length, $h_s$ and requires
\begin{equation}
\label{eq:grav_dt}
\mathrm{dt}_{i} = \sqrt{\frac{2 \eta h_{s,i}}{|\vec{a}|}},
\end{equation} 
where $\vec{a}$ is the total acceleration of the particle, $h_{s,i}$ is the smoothing length of the force at time step $i$, and $\eta$ is a tolerance parameter that determined what fraction of the softening length the particle is permitted to travel over the current time step. In traditional SIDM, the only inter-particle force is gravity so that $\mathrm{dt}_{i}$ is constrained by the particle separation and the time step decreases sharply as density increases\footnote{If the force softening is adaptive, and tied to the kernel size of the particle, then this will relax the time step for more diffuse regions, and shrink it with particle separation. }. 

The second criterion is set by $\kappa$, which controls the likelihood that an SIDM scattering events occurs during the timestep. Ideally we would want this number to be small to avoid scenarios in which multiple scattering events would be expected. In GIZMO this is implemented by reducing the local time step by $\kappa/\mathrm{prob}_{\mathrm{SIDM}}$ if $\mathrm{prob}_{\mathrm{SIDM}} > \kappa$. While adjusting the time-step allows the standard SIDM numerical approach to work over a wider range of densities at the cost of increased computing time, at some point the discretizations of the gas as macroscopic N-body particles will still be insufficient. The assumption that the simulation particle dynamics accurately tracks the micro-physical scattering breaks down when the mean free path of the physical particles is much smaller than the size of the simulation particles. That is, in the central region of the halo there can eventually be so many scattering events within the cell volume, during a single time step, that a prescription for the internal state of the simulation particle provides a more efficient and accurate description. This is the domain of hydrodynamics, which standard SIDM cannot approach, regardless of time-step size, and the reason the hybrid approach can do better.

There are additional challenges related to the time-stepping procedure found in SIDM simulations modeling gravothermal collapse, not readily alleviated by existing time stepping criteria \citep{Fischer:2024eaz, Mace:2024uze, Palubski:2024ibb, Zhong2023, Yang2022}. These largely relate to SIDM simulations in the frequent scattering regime exhibiting poor total energy conservation due to scattering events breaking the time-symmetry of the numerical method and the non-reversibility of the time-step changes. We again can avoid some of these by transitioning out of the SIDM method in the high frequency scattering regime, though SPH hydro-dynamics will of course introduce their own numerical shortcomings in its stead, and it doesn't remove the energy conservation defects SIDM introduces during the early stages of collapse. 

\begin{figure}
	\includegraphics[width=\columnwidth]{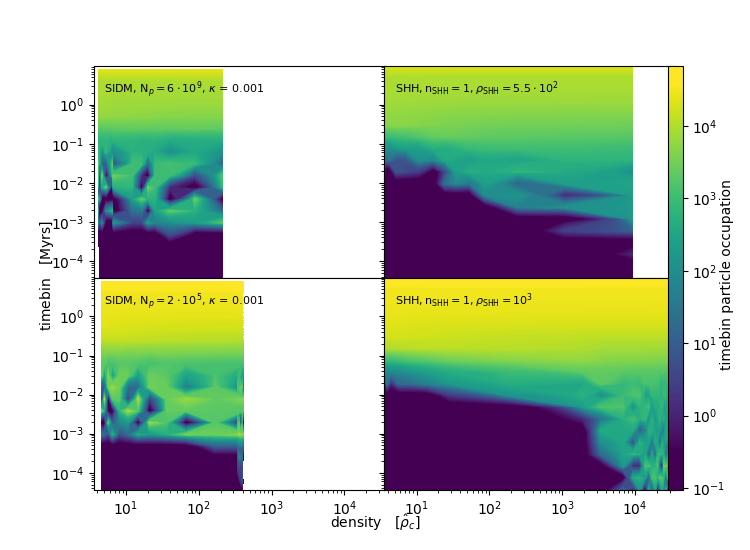}
    \caption{The difference in time bin occupation number in standard SIDM simulations (left) compared to SHH simulations (right) as a function of core density. The y-axis shows the length of the simulation time bins in Myrs and the x-axis the normalized density of the core. The color indicates the number of simulation particles in each time bin. (We assigned every empty bin $0.1$ of a particle for the sake of readability of the plot, so that dark indigo regions should be interpreted as empty bins.) The plot can be read as time bin occupation evolving with core density in time from left to right. The left most vertical slice shows the time bins at $\hat{\rho}_{c, min}$ and subsequent slices to its right the change in time bin occupation as the core density monotonically increases. The top (bottom) two panels show the lowest (highest) particle resolution $N_p=6 \cdot 10^4$ ($N_p=2 \cdot 10^5$). The high resolution halo has $\mathrm{n}_{\mathrm{SHH}} = 1$ and $\rho_{\mathrm{SHH}} = 10^{3}$, and its lower resolution counterpart the corresponding re-scaled transition density $\rho_{\mathrm{SHH}} = 5.5 \cdot 10^{2}$. In all cases $\kappa = 0.001$. \label{fig:SHH_SIDM_timebin}}
\end{figure}

In SHH, both time-step criteria undergo modification, with the fortunate byproduct of alleviating some of the SIDM slow down in high-density regions without detrimental impact on numerical accuracy. We first re-examine the acceleration term and note that since in our method all SIDM particles have gas properties, the total acceleration is now a product of the gravitational and the modified hydro components,
\begin{equation}
\label{eq:sph_dt}
dt_{i} = \sqrt{\frac{2 \eta h_{s,i}}{|\vec{a}_{g} + h(\rho)\vec{a}_{h}|}}.
\end{equation}

The contribution from the modified hydro acceleration term will generally oppose (via pressure) the gravitational one during core collapse (or close fly-by particle encounters). This results in the relaxation of the time step, compared to its SIDM-only counterpart, during the early evolution of the halo. Since in our method gravitation, hydro, and SIDM processes all share the same kernel smoothing length $h_{s,i}$, we set $\eta \leq \kappa$ so that once the core has collapsed, the time step size is primarily driven by the hydro terms. We no longer rely on $\kappa \rightarrow 0$ as $\mathrm{Kn} \rightarrow 0$ to control SIDM scattering since the particle dynamics follow the fluid equations in that regime which naturally take into account the internal energy of the particle.

We illustrate this in Figure \ref{fig:SHH_SIDM_timebin}, which shows the particle occupation-number of the time bins with increasing core density. Vertical slices from left to right show the change in time-bin occupation as the core density increases monotonically from $\hat{\rho}_{c, min}$. The plots on the left show SIDM only halos, with the new hybrid method on the right. The upper and lower rows are low- and high-resolution halos respectively. We note that as expected the small time bin occupation at the density nadir for SIDM halos is already considerably greater than for their hybrid method counterparts. The minimal time bin is also first occupied at considerably lower core densities in the SIDM case. Referencing Fig.\ref{fig:SHH_phase_time} which shows the evolution of the halo shown in the lower right of the time bin plot, we note that the density at which the hydro force terms become comparable to the gravitational terms, roughly corresponds to the sharp drop into smaller time bins set by the hydro time step criteria.

\section{Simulation results}
\label{sec:sims}
To compare the results of SHH to the standard SIDM simulations, we used \texttt{SpherIC} \citep{GarrisonKimmel2013}\footnote{https://bitbucket.org/migroch/spheric/src/main/} to generate initial conditions as Navarro-Frenk-White-type (NFW) \citep{Navarro1996,Zhao1995} isolated, spherically-symmetric halos. Such halos have a density profile of the form from Eq.(\ref{eq:nfw_profile}) with an additional cut-off term \citep{Zemp2008}:
\begin{equation}
    \rho(r)=\begin{cases}
    \rho_{\rm NFW}(r) & r\le r_{\rm cutoff}\\
    \rho_{\rm NFW}(r_{\rm cutoff})\left(\frac{r}{r_{\rm cutoff}}\right)^{\delta}e^{-\frac{r-r_{\rm cutoff}}{0.3 \; r_{\rm cutoff}}} & r>r_{\rm cutoff}
    \end{cases}.
\end{equation}
We used parameters following the high concentration halo from \citet{Palubski:2024ibb}, i.e. a $M_{200}=1.15\times10^9\,M_{\odot}$ halo with virial radius $R_{200}=22.1$ kpc, scale radius $r_s=0.715$ kpc, and a truncation radius $r_{\rm cutoff}=23.6$ kpc. This corresponds to a scale density $\rho_s=1.04\times10^{-1} M_{\odot}/$kpc and concentration $c_{200}=31$. The formula for $\delta$ is given in \citet{Zemp2008}; for the values listed here, $\delta=0.364$. 

To demonstrate how our new technique compares to others, we consider SIDM with a constant cross-section $\sigma_{m}\equiv \frac{\sigma}{m}=50\, {\rm cm}^2/{\rm g}$. This value is useful for illuminating the effects of self-interaction, although it is in tension with observational constraints which impose $\sigma_{\rm elastic}(v)/m_{\mathrm{DM}}\equiv\sigma_{m}(v)\lesssim\mathcal{O}\left[\left(10\,{\rm cm}^2{\rm /g}\right)\left(\frac{10 {\rm km/s}}{v}\right)\right]$ \citep{Sagunski:2021,Andrade:2021,Harvey:2019} (but see also \citet{Silverman:2022,Nadler2023}).

Parameters for all simulations, and the Figures the results appear in, in are shown in Table \ref{table:ICs}. There SIDM 
and SHH refer to the traditional SIDM and our new method, respectively. $N_{p}$ is the total
number of simulation particles, $\kappa$ is the interaction probability time-stepping criteria for SIDM simulations, and $\rho_{\mathrm{SHH}}$ and $n_{\mathrm{SHH}}$ are the bridging function parameter choices for SHH (see Eq.(\ref{eq:fKn})). We use $N_p = 6\times10^4$, $9\times10^4$, and $2\times10^5$ for convergence tests. 

\begin{table*}
\centering
\begin{tabular}{l | c | c | c | c | c | c | }
\hline\hline
     Kinetic prescription & $N_{p}$  & $\kappa$ & $n_{\mathrm{SHH}}$ & $\rho_{\mathrm{SHH}}$  & $h(\rho|_{\mathrm{Kn_{sim}} = 10^{-3}})$ & Figure\\ 
      &  & & & & &  $[\mathrm{M}_{\odot}/ \mathrm{pc}]$ \\ 
     \hline

     SIDM & $[6, 9] \times 10^{4}$ &  0.001& -  &  - & - &Fig. \ref{fig:Collapse_SIDM}  \\      
     SIDM & $2 \times 10^{5}$ & 0.02, 0.002& -  &  - & - &Fig. \ref{fig:Collapse_SIDM}  \\
     SIDM &  $2 \times 10^{5}$ &  0.001& -  &  - & - & Fig. \ref{fig:Collapse_SIDM}, \ref{fig:SHH_function} \\ 
     \hline
     
     SHH &  $2 \times 10^{5}$ &  0.02, 0.002&   1  & $1 \times 10^{3}$ & $0.993$ &Fig. \ref{fig:SHH_configure}  \\  
     SHH & $2 \times 10^{5}$ &  0.001 &  1  &$1 \times 10^{3}$ &  $0.993$ & Fig. \ref{fig:SHH_configure},  \ref{fig:SHH_function},  \ref{fig:core_profiles}, \ref{fig:SHH_phase_time}, \ref{fig:SHH_phase_function}\\ 
     SHH & $9 \times 10^{4}$ &  0.001 &  1 &$6.7 \times 10^{2}$ & $0.993$ & Fig. \ref{fig:SHH_configure} \\   
     SHH & $6 \times 10^{4}$ &   0.001&   1 & $5.5 \times 10^{2}$ & $0.993$ &  Fig. \ref{fig:SHH_configure}  \\          
     \hline
      SHH &  $2 \times 10^{5}$ & 0.001& 2/3 &  $[5 \times 10^{3}, 1 \times 10^{4}]$ & [$0.964$, $0.854$] &  Fig. \ref{fig:SHH_function}  \\  
      SHH &  $2 \times 10^{5}$ & 0.001 &  1 &$[5 \times 10^{2}, 1 \times 10^{4}]$ & [$0.964$, $0.854$] &  Fig. \ref{fig:SHH_function}  \\  
      SHH &  $2 \times 10^{5}$ & [0.02, 0.0002] &  1 &$ 1 \times 10^{4}$ & 0.854 &  Fig. \ref{fig:SHH_function}  \\        
     SHH & $2 \times 10^{5}$ &  0.001&  3/2 &$1 \times 10^{3}$ &  $0.999$ & Fig. \ref{fig:SHH_function}   \\  
\hline \hline
\end{tabular}
\caption{Initial conditions and simulation parameters used in this work. The first column (kinetic prescription) indicates whether the halo was evolved using the standard SIDM algorithm or our proposed SHH method, $N_{p}$ is the particle resolution of the halo, $\kappa$ is the scattering probability tolerance, and lastly $r_{SHH}$ and $n_{SHH}$ are the free parameters of the matching function used in our proposed method. All simulations were run starting at a redshift $z_{c}=31$, with gravitational time-stepping tolerance, $\eta=0.001$ (Section \ref{sec:timestepping}) and NFW scale radius of $r_s=0.715$ kpc and scale density $\rho_s=0.0104$ $\mathrm{M}_{\odot}/\mathrm{pc}^3$ (Section \ref{sec:sims}). Likewise, the gravitational, self-interacting, and hydrodynamical force softenings were all set to adaptive (Section \ref{sec:GIZMO_mods}). \label{table:ICs}}
\end{table*}

\subsubsection{Central density and energy conservation calculation }
To compare our results to those of the results of the semi-analytic prescriptions shown in Fig. \ref{fig:reviewSIDMcurves}, we need to define the central density. For our simulated halos, we derive this directly from the kernel-constructed density around each particle in GIZMO. This method is in contrast to the density profile fit of the core region of the halos performed in, for example, \cite{Palubski:2024ibb}. For the gas particles, this is done trivially as part of the hydro loop, while for the N-body particles in the SIDM only simulations, the SIDM scattering loop constructs a local density when calculating the scattering probability. We select the particle with the lowest potential (also calculated with GIZMO on the fly) and sample a resolution dependent number of nearest neighbors to arrive at a core density. The number of nearest neighbors is chosen so that core densities reproduce those produced with the profile fitting method used in \cite{Palubski:2024ibb}. 

We track the total energy of the halo over the time of the simulation via the total gravitational potential, kinetic, and internal energy of all the particles calculated by GIZMO at specified time intervals. The internal energy is an inherent attribute of the SPH gas particles in GIZMO, and updated along with the momentum kick each time step according to Eq.~(\ref{eq:SecMoment}). In the case of SIDM only simulations, the internal energy component is set to zero. The error in the simulations is presented as a fraction $(E_{\rm grav} + E_{\rm kin} + E_{\rm int})/|E_{0}|$ in the following plots.

The frequency at which GIZMO collects energy statistics does not always align with the simulation time steps of the particles. This can become noticeable in cases where dynamic quantities are only updated at the end the respective timestep. For ease of readability we have chosen the statistical sampling interval to be frequent enough relative to the simulation snapshots intervals to allow for a rolling average to be calculated. 

\subsection{SIDM only results}
As explored in recent work \citep{Meskhidze:2022,Mace:2024uze,Palubski:2024ibb}, the collapse time of the SIDM halo is highly sensitive to not just the dynamic equations used, but also the numerical configuration of the code in question. This makes assessing the veracity of a new method challenging, as there is no benchmark collapse time to compare with. For reference, we begin by generating collapse curves using the traditional SIDM method, with different simulation parameters.

We ran SIDM halos with parameter choices very similar to those presented in \cite{Palubski:2024ibb}, listed at the beginning of the section. The Spheric N-body particle initial conditions are directly comparable to their high-redshift-collapse halos and used directly as generated. Halos with low ($6 \times 10^{4}$), medium ($9 \times 10^4$), and high ($2 \times 10^{5}$) particle resolution are shown together with different SIDM scattering probability criteria $\kappa$ (see Table \ref{table:ICs}). We used a fixed $\eta$ of $0.001$ throughout our work, which is lower than that used in \cite{Palubski:2024ibb}, to accommodate the use of the hydro loop in later results. 

\begin{figure}
	\includegraphics[width=\columnwidth]{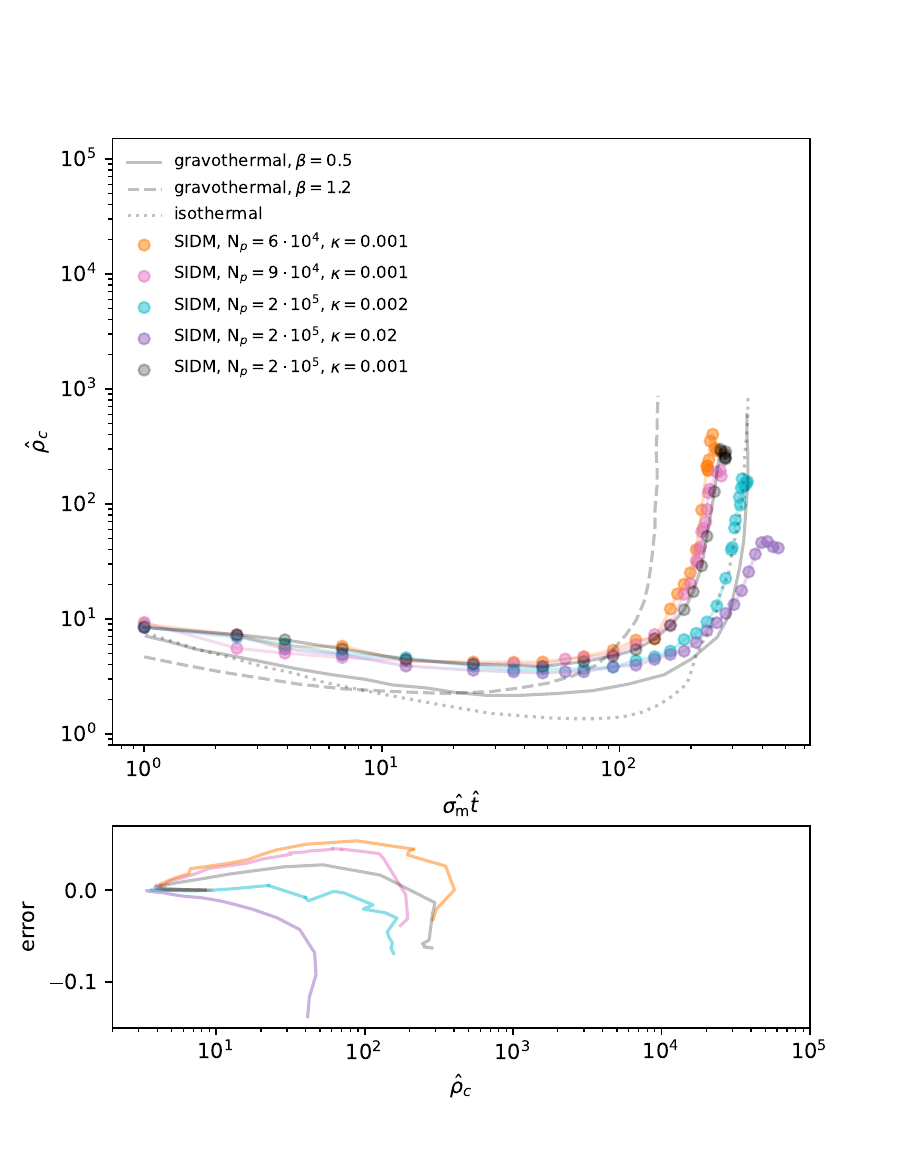}
    \caption{Core collapse with SIDM, using different simulation configurations (see Table\ref{table:ICs}), showing the rescaled density against rescaled time. The semi-analytic solutions for the gravo- and iso-thermal collapse are also shown for reference in gray-scale. The lower panel shows the violation of energy conservation, as the fractional error $\delta E/|E_{0}|$ in the total energy, accrued over the simulation, plotted against the re-scaled density using the colors corresponding to the curves and legend in the top panel.  \label{fig:Collapse_SIDM}}
\end{figure}

Results are shown in Fig. \ref{fig:Collapse_SIDM}. The upper panel shows core-collapse curves with the density and time axes normalized as described near Eq.(\ref{eq:nfw_profile}). Semi-analytic results from \cite{Yang2023} for both the gravothermal and isothermal methods are shown in gray for reference. Orange, pink, and gray circles show halos with increasing particle resolution (low, medium, high) and fixed $\kappa = 0.001$, while cyan and purple show high-resolution halos with larger $\kappa$ (0.002, 0.02, respectively). As reported in other works, we note that as the central densities reach $\approx 10^2 \hat{\rho}_{c}$, collapse either stalls (black, pink, cyan curves) or outright reverses (purple curve). Curves with a greater $\kappa$ collapse at later times, while those with $\kappa = 0.001$ appear to favor earlier collapse. All our SIDM solutions display a shallower core density dip towards $\hat{\rho}_{c, \mathrm{min}}$ compared to some of the results in the literature. This may be due to the morphology of the halo generated by our particular choice of Spheric parameters, or additional configuration parameters used in our build of GIZMO.

The lower panel of Figure \ref{fig:Collapse_SIDM} shows the fractional error in total simulation energy conservation as a function of core density and sheds further light on the collapse behavior. The large $\kappa$ simulation loses energy from very early on the in the halo's evolution, which likely informs their late collapse time. In contrast, those with $\kappa = 0.001$ initially gain energy, which reverses as the central density increases. The initial energy gain may be due to numerical heating, caused by the adaptive force softening and small choice of $\kappa$. We find that for this particular scenario, increasing the resolution tends to lessen the error accrued by the simulation. Reducing the SIDM scattering tolerance improves energy conservation up to a point. Regions in which the core density either stalls or reverses are associated with considerable loss in total simulation energy. Judicious choice of configuration parameters can reduce these errors, though at commensurate computational cost. Other aspects of the SIDM method, such as the time symmetry breaking of the scattering events, remain as contributors of error in high scattering domains.

\subsection{SHH results}
\begin{figure}
	\includegraphics[width=\columnwidth]{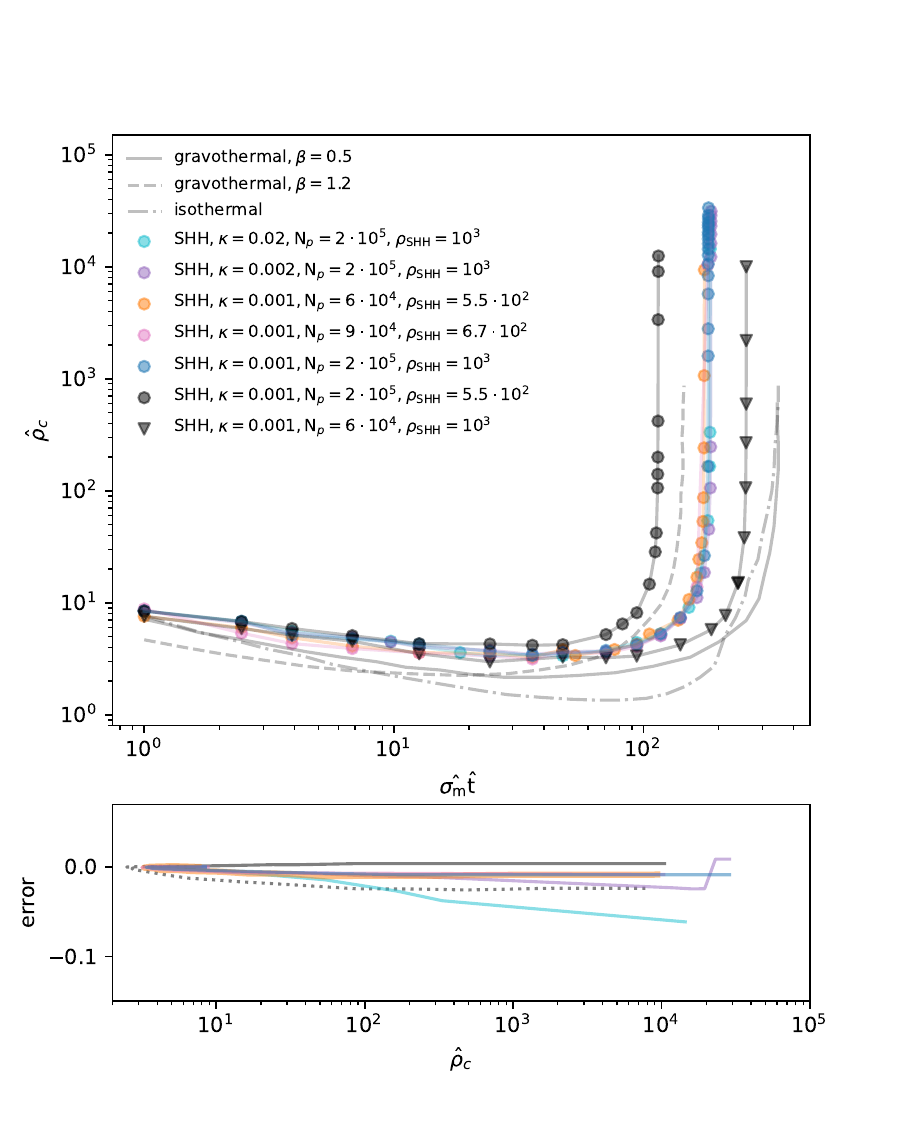}
    \caption{Core collapse with SHH, using different simulation configurations (see Table\ref{table:ICs}). The cyan, purple, orange, pink, and blue curves show that the simulation is robust to changes in the SIDM scattering parameter $\kappa$ and has good convergence with increasing resolution as long as $\rho_{\rm SHH}$ is consistently determined using the simulation Knudsen number, Eq.(\ref{eq:ComputerhoSHH}). The gray circles and triangles show the effect of changing particle resolution with adjusting the Knudsen number accordingly. The semi-analytic solutions for the gravo- and iso-thermal collapse are also shown for reference. The lower panel shows the violation of energy conservation, demonstrating that the SHH technique accrues significantly less error out to higher densities compared to SIDM. The orange line width in the lower plot has been increased relative to the other curves for the sake of readability.  \label{fig:SHH_configure}}
\end{figure}

Next, we perform an equivalent exploration of the simulation parameter space for the SHH method. For details please refer to the SHH blocks in Table \ref{table:ICs}. As was done for SIDM in Figure \ref{fig:Collapse_SIDM}, we compare the impact that changing the scattering parameter $\kappa$ and the particle resolution has on the collapse behavior of the halo. Results are shown in Figure \ref{fig:SHH_configure}. For all simulations in this Figure, $\mathrm{n}_{\mathrm{SHH}} = 1$. The cyan, purple, and blue curves correspond to a high-resolution halo, with $\rho_{\mathrm{SHH}} = 10^{3}$, but with decreasing $\kappa$. The orange and pink curves in Figure \ref{fig:SHH_configure} show lower resolution halos, all with $\kappa = 0.001$ and For these curves $\rho_{\mathrm{SHH}}$ was adjusted to compensate for the decrease in resolution according to Eq.(\ref{eq:ComputerhoSHH})). (Fixing $\rho_{\mathrm{SHH}}$ to be determined ${\rm Kn_{sim}}=10^{-3}$ means that $\rho_{\mathrm{SHH}}$ decreases when particle resolution decreases, i.e. simulation particle mass goes up). For comparison, we show both the high- and low-resolution halos without the $\rho_{\mathrm{SHH}}$ adjustment for the simulation Knudsen number in the black curves. The reference curves for the semi-analytic models are again included in gray.

We note that for the new method there is greater consensus in collapse time, regardless of choice of $\kappa$, and all three halos commence collapse well beyond $10^3\hat{\rho}_{c}$ without stalling or reversal. Halos with different particle resolution also appear to converge, although the particle placement in the respective initial conditions files will here account for some of the difference in collapse time. The two reference curves in black indicating models for which the simulation particle resolution is not accounted for in the bridging function, show a marked difference in collapse time. This confirms the importance of the cell Knudsen number in numerical methods, in addition to its physical counterpart. 

The lower panel of Figure \ref{fig:SHH_configure} shows the accumulation error for our method, with the same axis dimensions as Fig.\ref{fig:Collapse_SIDM} for ease of comparison. Perhaps the most striking feature is that for our equivalent models with $\kappa = 0.001$, halos display considerable less erroneous gain in energy than their SIDM counterparts. We may speculate that the presence of the hydro force term modifies the dynamics of close particle encounters and could feasibly lead to some suppression of numerical heating. This bears further investigation in future work. We note that while there is some loss of energy in all simulations as the halo undergoes gravothermal collapse, it does not appear to be as drastic as in the SIDM-only case and this robustness may be a reflection of the core's continued ability to collapse. The two most aberrant curves are for $\kappa = 0.02$ and $0.002$, with the former's continued loss extending well beyond that of the other simulations. The sharp increase found towards the high density end of the purple curve is likely from the SPH solver's well-documented numerical shortcomings in collapsing gas environments, exasperated in this scenario by our small choice in nearest neighbor number \citep{Price2012, Read2010}\footnote{Adaptive Mesh Refinement methods exist which combine desirable aspects of SPH and Riemmann solvers that are more adept at dealing with shock fronts and discontinuities. Unfortunately, the way these treat momentum conservation across cell boundaries makes them incompatible with this hybrid methods.}. 

\begin{figure}
	\includegraphics[width=\columnwidth]{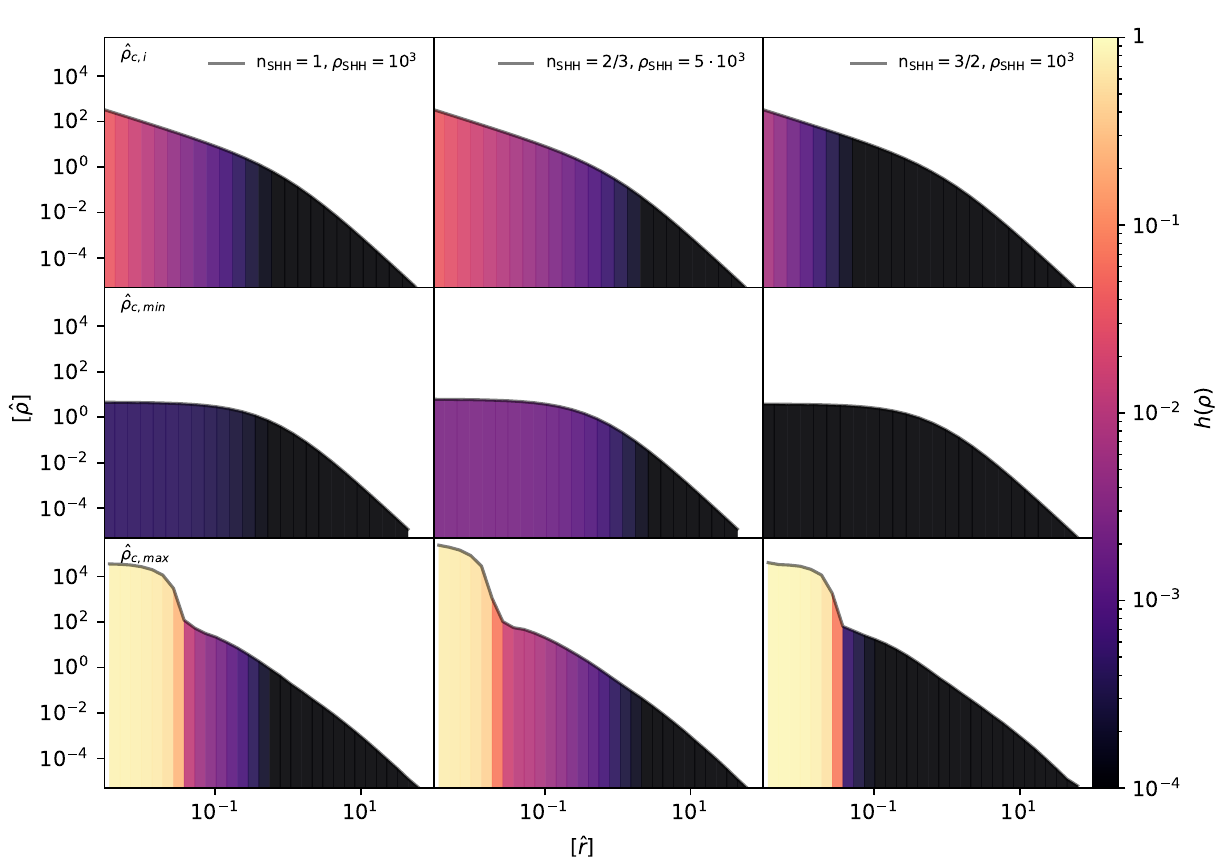}
    \caption{The value of the bridging function, $h(\rho)$, as a function of density across a radial slice of the halo. The vertical axis shows density while the color bar indicates the size of $h(\rho)$ (Eq.(\ref{eq:fKn})). From left to right, columns show results for $\mathrm{n}_{\mathrm{SHH}} = 1, 2/3$, and $3/2$ and  $\rho_{\mathrm{SHH}} = 10^3, 5 \cdot 10^3$, and $10^3$ respectively. Within each column, three different times are shown: the simulation start (top), the snapshot when $\hat{\rho}_{c, min}$ (middle), and the final snapshot after the core has formed (bottom). \label{fig:Kn_funct_pcre}}
\end{figure}

As discussed near Eq.~(\ref{eq:ComputerhoSHH}), the local Knudsen number informs the bridging function but some numerical ambiguity as to the correct choice of fitting parameters remains. Next, we demonstrate the impact of changing the onset and steepness of the transition between standard SIDM and hydrodynamical equations. The parameters considered are listed in Table \ref{table:ICs}, and the corresponding bridging functions were shown in Figure \ref{fig:Knudsen_functions}. In Figure \ref{fig:Kn_funct_pcre} we show the value of the bridging function achieved in the simulations for three parameter sets, each at three different time snapshots. The Figure shows that most of the kinetic solution is nearly SIDM regardless of parameter choice. However, the bridging functional form does substantially change how often there is a percent level hydrodynamical component. The region where $h(\rho)\sim\mathcal{O}(0.1)$ remains small in all cases. Nonetheless, as we will see in the next figure, those small contributions do affect the overall collapse dynamics. For all parameter choices, the inner core is firmly in the hydrodynamical regime by the end of the simulation.

\begin{figure}
	\includegraphics[width=\columnwidth]{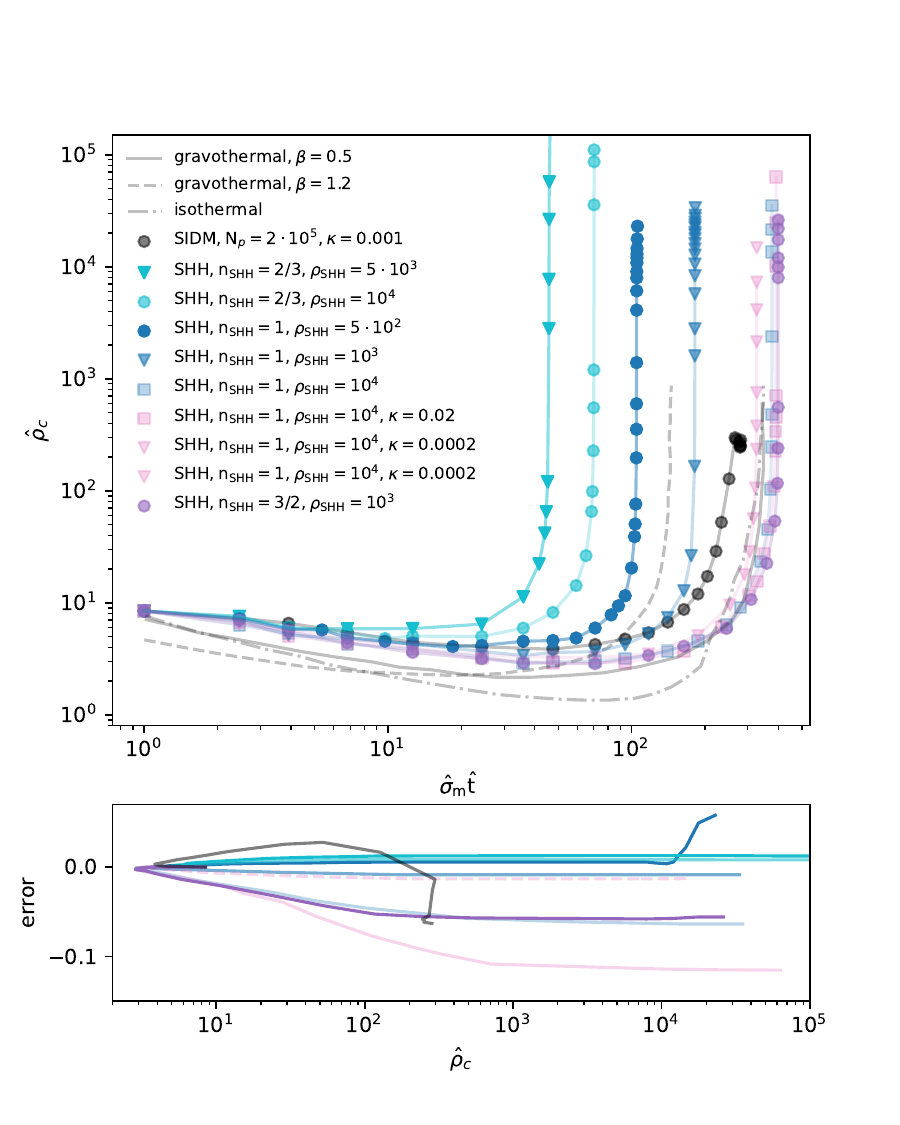}
    \caption{Core collapse using the SHH method, for different values of the parameters in the bridging function, $h(\rho)$ as given by Eq.(\ref{eq:fKn}). All runs use $2 \times 10^{5}$ simulation particles and $\kappa = 0.001$. (Refer to Figure \ref{fig:Knudsen_functions} for the variation in the function $h(\rho)$ for this set of parameter choices, and to Figure \ref{fig:Kn_funct_pcre} to see how $h(\rho)$ varies across the simulated halos at different time steps.) The lower panel shows the violation of energy conservation. 
     \label{fig:SHH_function}}
\end{figure}

Figure \ref{fig:SHH_function} shows the simulation results, including a benchmark standard SIDM simulation at high-resolution and $\kappa=0.001$ (in gray) for comparison. The Figure shows results for $\mathrm{n}_{\mathrm{SHH}} = 1, 2/3$, and $3/2$, and with different values of $\rho_{\mathrm{SHH}}$, set by changing the critical value of ${\rm Kn_{sim}}$ used in Eq.~(\ref{eq:ComputerhoSHH}) (see Table\ref{table:ICs}). As expected, a bridging function with lower $\rho_{\mathrm{SHH}}$ and shallower index $n_{\mathrm{SHH}}$ will produce an earlier collapse time compared to the benchmark SIDM model. What is perhaps more interesting is the two curves implementing a high $\rho_{\mathrm{SHH}}$ threshold coupled with a high $n_{\mathrm{SHH}}$ index will produce collapse times later than the SIDM example in gray. This may be driven by the hydro forces becoming relevant at later stages in the collapse, so that the energy loss encountered during SIDM scattering is more pronounced (akin to the $\kappa = 0.02$ and $0.002$ SIDM only models in Figure~\ref{fig:Collapse_SIDM}). The reduced hydro term is still sufficiently present to suppress the numerical heating encountered in low $\kappa$, adaptive softening models, so the energy loss during the SIDM dominated phase leads to delayed collapse compared to the equivalent $\kappa = 0.001$ SIDM only halo.  

As Figure \ref{fig:SHH_function} shows, the choice of bridging function can lead to a wide range of collapse times, in a similar manner as the choice of $\beta$ in the gravothermal models. Just as work on the gravothermal prescription for SIDM has led to an understanding about the range of $\beta$ that is likely sensible, further semi-analytic arguments of SHH simulations of particular halo initial conditions will likely narrow the range of reasonable parameter choices compared to the scope shown in Figure \ref{fig:SHH_function}. Looking back to Figure \ref{fig:Knudsen_functions}, we note that curves which produce very early collapse times still have a contribution from the hydro terms well into regimes where the Knudsen number is bigger than 1. This may indicate that the suppression of the hydro term at low densities is as important to consider as the high-density transition into the Euler solutions when setting the bridging function. 

In contrast, the models that produce the latest collapse times are heavily suppressed in regions where the scattering mean free path just begins to drop below the cell scale. (Compare the cyan curves having a lower contribution than these curves at high densities, but still producing much earlier collapse times due to the contribution at low densities.) These models do, however, provide a good match for the iso-thermal semi-analytic model from \cite{Yang2023}, and the gravothermal model with weak $\beta$ coupling. This is perhaps not surprising, as the iso-thermal model makes use of SIDM solutions in the outer parts of the halo, and the low $\beta$ parameter leads to a sharp transition between the low and high frequency scattering domains, so all essentially fall into a category of having a sharp, high density switch to fluid dynamic physics.

We again note that our models do not exhibit the sharp increase in total simulation energy compared to their SIDM only counterparts, despite setting $\kappa = 0.001$. The two late collapsing curves however, show considerable loss in total simulation energy throughout the halo's early evolution, though this halts as the hydrodynamics takes over, suggesting this numerical error is still driven by SIDM. This is further illustrated by the pink curves, showing the $\mathrm{n}_{\mathrm{SHH}} = 1$, $\rho_{\mathrm{SHH}} = 10^4$ model with $\kappa = 0.02$ and $0.0002$. Relaxing $\kappa$ to $0.02$ shows stark loss of total simulation energy conservation, and delays collapse compared to the equivalent model with $\kappa=0.001$. In contrast reducing $\kappa$ to $0.0002$ achieves energy conservation comparable to the $\rho_{\mathrm{SHH}} = 10^4$ model and leads to appreciable earlier collapse. The $\rho_{\mathrm{SHH}} = 5\cdot 10^3$, $\mathrm{n}_{\mathrm{SHH}}=1$ curve in dark blue also seems to show the steep energy gain indicative of the SPH solver encountering a sharp increase in pressure. Overall it is encouraging to note that our method produces simulations in which the energy conservation error is either comparable to SIDM methods, or produces more favorable results.

Our Goldilocks candidate is the model with $n_{\rm SHH}=1$ and $\rho_{\mathrm{SHH}}=10^3$ (blue curve with triangle markers). It cuts off sharply for Knudsen numbers $\leq1$ and provides a substantial onset of hydro forces for $\mathrm{Kn} \leq 10^{-2}$. It also follows the benchmark SIDM curve during the turn-around phase of the evolution until proceeding to full collapse without stalling. While not intended to be taken as any kind of "true'' solution, we shall use this model to illustrate some additional features of the collapsed gravothermal core found in SHH models.

Lastly, we want to briefly discuss the role that both halo morphology and environment may play in the collapse behavior of SIDM halos. Previous works \citep{Palubski:2024ibb,Gad-Nasr:2023gvf} have shown that the shallower cusps formed in halos collapsing at lower redshifts certainly lend themselves to longer core collapse times. In some preliminary work, we also found evidence that in simulations where the initial conditions traced the halo's evolution from an initial over-density rather than an already formed NFW profile, collapse times were also earlier. We speculate that this may be due to the density spike produced during the initial collapse of the over-density into the halo after turn-around. Similarly, the impact of angular momentum and the merger history of a halo may turn out to be of interest when setting a characteristic core collapse time for a fixed SIDM cross-section. The interconnected effect of SIDM and morphology may have particularly significant at high redshift and subsequently impact the first stars and galaxies hosted by these early collapsed structures. We hope to follow these lines of inquiry in future work, as well as show how these factors manifest in cases where the dark-matter scattering cross section is velocity suppressed.

\subsection{Collapse features}
\begin{figure}
	\includegraphics[width=\columnwidth]{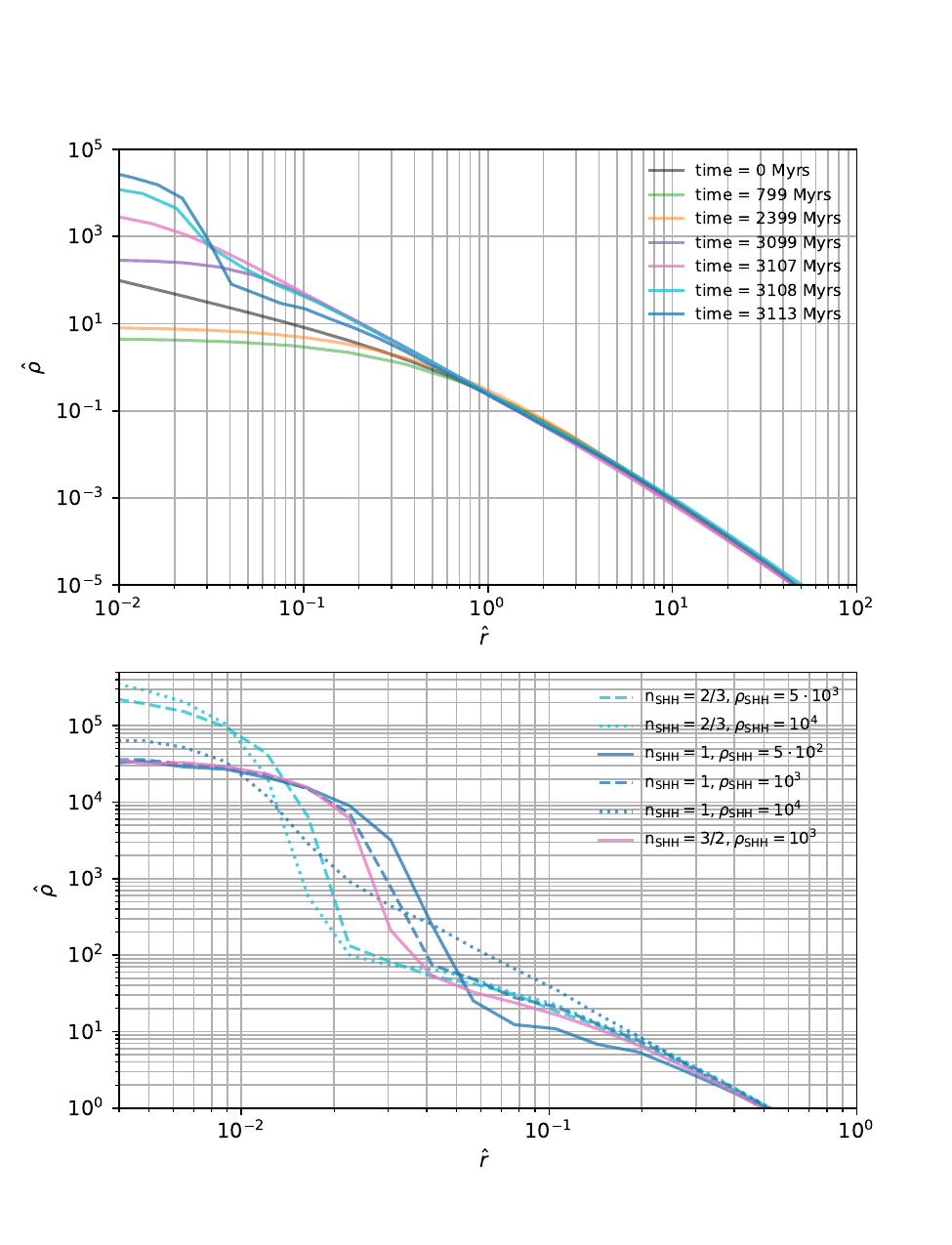}
    \caption{The upper plot shows the evolution of the core density profile of a $N_p=2 \cdot 10^{5}$ particle halo with $\mathrm{n}_{\mathrm{SHH}} = 1$, and $\rho_{SHH} = 10^3$ halo over time. The lower plot shows the core profiles obtained from high resolution halos with different bridging functions. Since the different models produce different collapse times, the profiles should not be taken as synchronous. The color and line style of the curves matches those of their corresponding bridging functions in Fig.  \ref{fig:Knudsen_functions}.  \label{fig:core_profiles}}   
\end{figure}

The distinguishing feature of our proposed method is its dynamic solution's dependence on the local Knudsen number shrinks, transitioning from traditional SIDM into a full fluid description at sufficiently high densities. The macroscopic prescription in this regime includes pressure as a manifestation of the local internal energy which provides a counter to the gravitational forces as the core begins to collapse in on itself. As a result of this, we note some distinctions in comparison to the central cores produced in previous methods.

Figure \ref{fig:core_profiles} shows the evolution of the core profile of our fiducial $N_p=2 \cdot 10^{5}$ particle halo with $\mathrm{n}_{\mathrm{SHH}} = 1$, and $\rho_{\rm SHH} = 10^3$ in the upper plot. The halo's core follows the behavior predicted by the semi-analytical methods, as shown in \cite{Yang2023} for example, in which the core initially hollows out before collapse commences. We note a similar flattening in the central profile up to around $\approx10^3\hat{\rho}_{c}$, before the final cusp forms. In the upper plot, $19565$ particles are used to resolve the core within a radius of $10^0 \, \hat{\mathrm{r}}$. (Comparable values hold for the lower Figure showing the core comparison for different bridging functions. The model with the steepest index $\mathrm{n}_{\mathrm{SHH}} = 3/2$ has somewhat fewer particles within that radius.) 

\begin{figure*}
	\includegraphics[width=\textwidth]{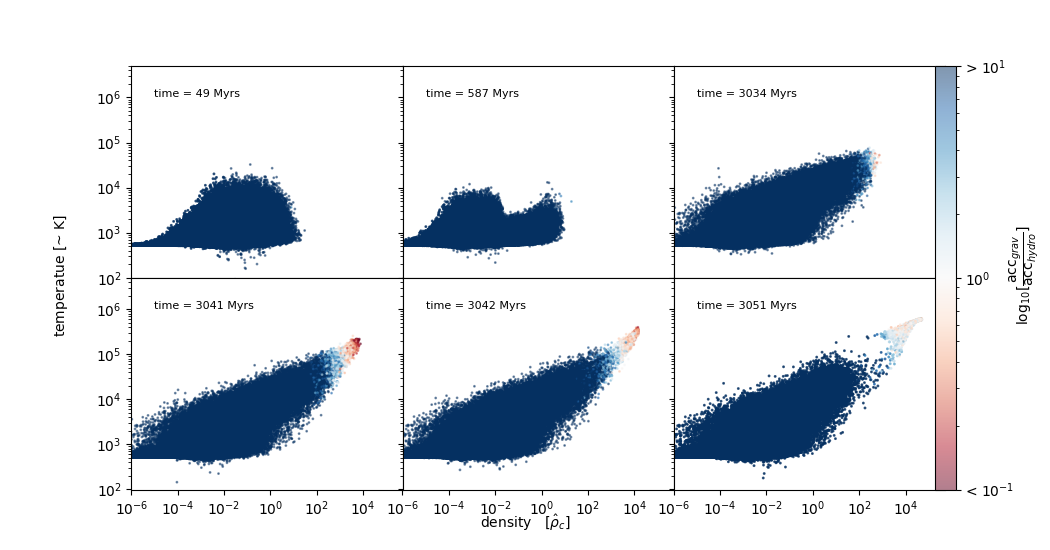}
    \caption{Phase plot of the collapsed core over time for the $\mathrm{n}_{\mathrm{SHH}} = 1$ and $\rho_{\mathrm{SHH}} = 10^{3}$ halo, with $2 \cdot 10^5$ particles. The color map shows the log scaled ratio between the magnitude of gravitational acceleration and the modified hydro acceleration.   \label{fig:SHH_phase_time}}
\end{figure*}

The lower plot shows the final snapshots of the dense core formed in high resolution halos with $\kappa =0.001$ for various bridging functions shown in Figure \ref{fig:Knudsen_functions}. We note the morphological difference between the cores formed, indicating the impact the bridging function plays beyond the collapse time of the halo. The two cyan curves with $\mathrm{n}_{\mathrm{SHH}}= 2/3$ produce denser cores which may be attributed to the hydro component contribution, and therefore pressure, being smaller than in other models at the same high densities. The dotted blue curve has lower contributions from the hydro forces relative to other models at both high and low densities and has the least steep change in density gradient approaching the central core. The remaining blue and pink curves have comparable high density hydro contributions, though vary in their respective low density fall off. They display similar core densities albeit with notable difference in their size. 

\begin{figure*}
	\includegraphics[width=\textwidth]{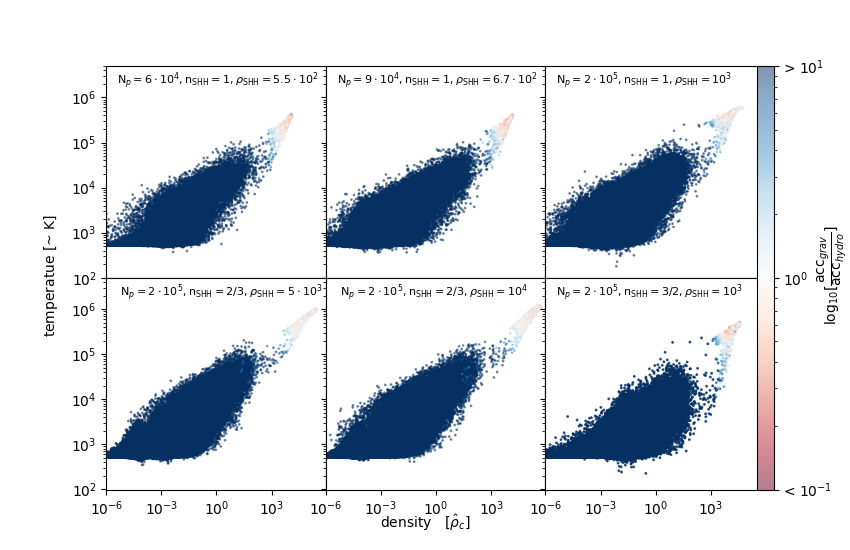}
    \caption{Same as Figure \ref{fig:SHH_phase_time}, but showing the latest phase plot in the evolution for different bridging functions. (Latest snapshots are not at the same time for different halos. Halo runs were terminated when the simulation's smallest time steps dropped below [check precise value].) The top row shows increasing resolution halo for the $\mathrm{n}_{\mathrm{SHH}} = 1$ and $\rho_{\mathrm{SHH}} = 10^{3}$ scenario, with resolution scaled parameters for the lower particle number halos. The lower row shows high resolution halos with different transition functions. The two right most have $\mathrm{n}_{\mathrm{SHH}} = 2/3$ with $\rho_{\mathrm{SHH}} = 5 \cdot 10^{3}$ and $10^4$ respectively, while the left most shows a model with $\mathrm{n}_{\mathrm{SHH}} = 3/2$ and $\rho_{\mathrm{SHH}} = 10^{3}$.    \label{fig:SHH_phase_function}}
\end{figure*}

We illustrate the relative contributions of hydro and gravitational forces in Figures \ref{fig:SHH_phase_time} and \ref{fig:SHH_phase_function}. In both Figures, each panel shows the density-temperature phase plot of the halo at the indicated snapshot time. The color bar shows the ratio of the particle's gravitational acceleration to that of the modified hydro acceleration term on a log scale. Dark blue is gravitationally dominated, while red is hydro dominated. (The scale is truncated at $10^{-1}$ and $10^{1}$ respectively for visual ease.) Figure \ref{fig:SHH_phase_time} shows the evolution of the fiducial $N_p=2 \cdot 10^{5}$, $\mathrm{n}_{\mathrm{SHH}} = 1$, $\rho_{\mathrm{SHH}}=10^3$ reference halo over time, while Figure \ref{fig:SHH_phase_function} shows the last time step of evolution for different particle resolutions (top panels) and different bridging functions (bottom panels).

The time-evolution in Figure \ref{fig:SHH_phase_time} shows that the pre-collapse phase is driven by gravitational forces, as expected. Over the next $\approx 3000$ Myrs, the inner part of the halo begins to become denser and hotter. This stage of the evolution is driven predominantly by SIDM scattering kicks, though the heavily suppressed hydro forces also play a role, especially during close encounters of the simulation particles. Eventually, the core density and temperature become sufficiently high for the hydro forces to fully come into play. A key shift in the dynamics arises from the hydro-facilitated conversion of kinetic bulk energy into internal energy, akin to the shocks undergone during star forming gas's collapse into a halo's potential well. This is shown in the bottom three panels of Fig. \ref{fig:SHH_phase_time}, where over a short time there is a rapid increase in the importance of hydro forces, followed by a local spike in temperature, before the central density settles into an apparent iso-thermal core. The balance between the hydro and gravitational acceleration at this point may indicate a halt to gravothermal collapse in the simulation, though since the time-step of our simulation at this point reached a bottle neck, we cannot make any declarations on the state of system beyond that point. Meanwhile, inclusion of the non-ideal thermal conductivity term would further alter the behavior of the halo at this point and lead to further collapse. We defer an extension to the present methodology to include non-ideal fluid equations terms for future work. 

Fig. \ref{fig:SHH_phase_function} again shows density-temperature phase-space plots, with the color scale indicating the ratio of gravitational and modified hydro forces. Here however, we show the final snapshot of halos with various different simulation and model parameters. The top row shows halos with different particle resolutions, culminating in our reference halo on the right. The previous two have model parameters ($\mathrm{n}_{\mathrm{SHH}} = 1$, and $N_p= 6\cdot 10^4$, $\rho=5.5 \cdot 10^2$ and $N_p=9 \cdot 10^4$, $\rho=6.7\cdot10^2$ respectively) to account for the change in simulation particle resolution within our model framework. The lower row shows three high-resolution, $N_p=2\cdot 10^5$ halos, with different $\mathrm{n}_{\mathrm{SHH}}$ parameters, differentiating between gradual and sharp onsets of the hydro-force component. The left and center plots show the former scenario $\mathrm{n}_{\mathrm{SHH}}= 2/3$, while the plot on the right shows the latter with a $\mathrm{n}_{\mathrm{SHH}} = 3/2$. The termination point of each simulation is somewhat arbitrary, as we let each halo run until a timestep bottleneck was reached, so the corresponding phase plots should not necessarily be taken as one-to-one comparisons, but instead a comparison of the qualitative behavior post-collapse.

All halos display balancing of hydro and gravitational forces as the gravothermal collapse core forms. The phase space tear between the inner and central part of the halo also appears to be ubiquitous. In addition to the difference on core morphology shown in Figure \ref{fig:core_profiles}, here we also observe the difference in core temperature for different bridging functions. While beyond the computational resources of this particular suite of simulations, it would be of interest to see whether the self-interacting cusp of the SIDM dominated halo reforms after the initial core collapse with time. Continued accretion would have interesting implications for the final size of the dense central object.  

\begin{figure*}
	\includegraphics[width=\textwidth]{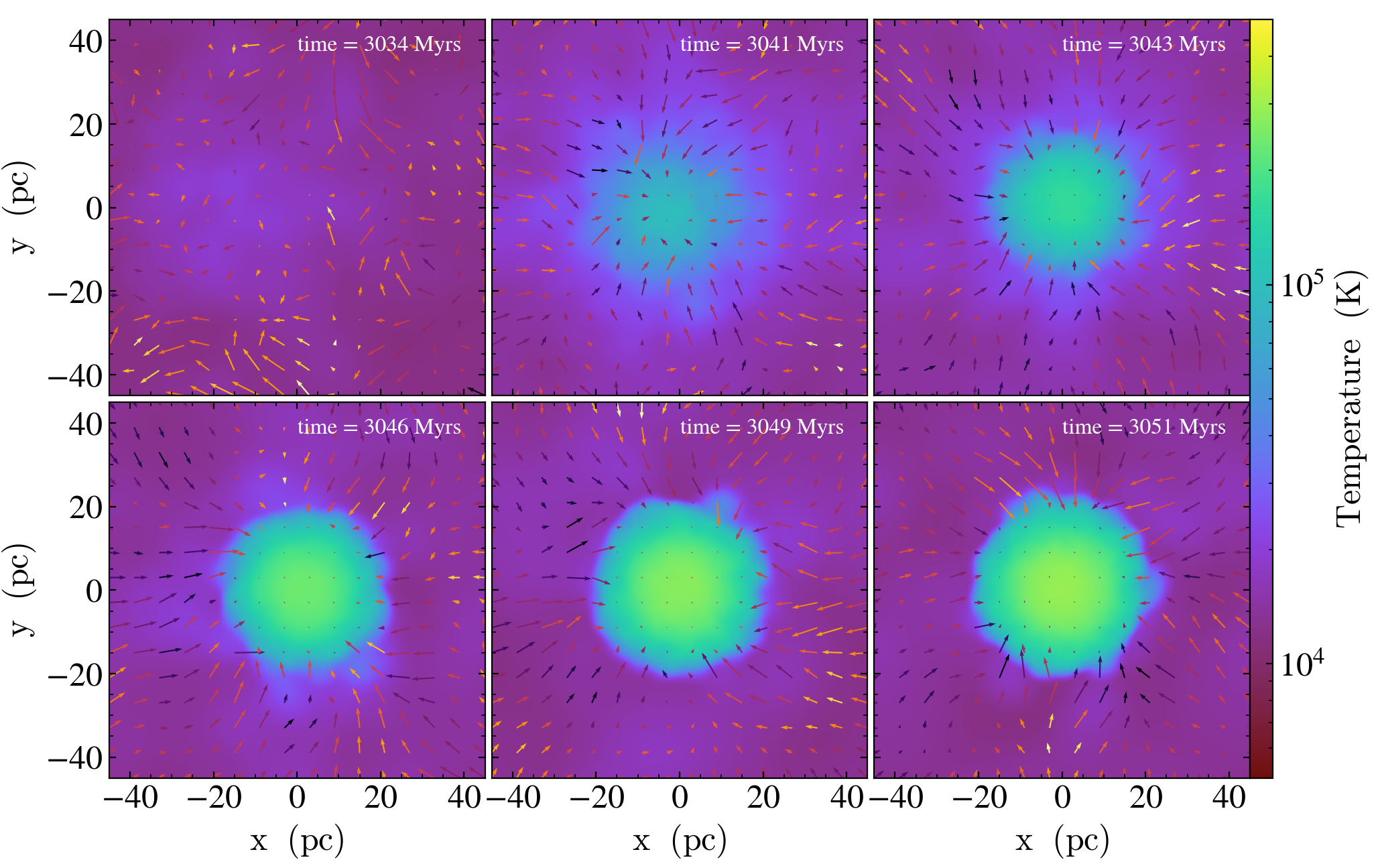}
    \caption{Temperature projection plot of the collapsing core over time, for a $10^5$ particle halo with $\mathrm{n}_{\mathrm{SHH}} = 1$ and $\rho_{\mathrm{SHH}} = 10^{3}$. The arrow overlay indicates the local gas velocity in the plane, with the color map indicating the perpendicular ($z$) component of the velocity field.}
    \label{fig:SHH_temperature}
\end{figure*}

Figure \ref{fig:SHH_temperature} provides an additional perspective on the dense post-collapse core. Each plots' origin is the center of mass of the system (there is some drift of the dense core during collapse relative of the rest of the halo). The color bar indicates temperature, and the arrows represent the underlying gas velocity field. The color of the arrow corresponds to the z-component of the velocity field. The time period covered here spans the same interval as that covered by panels 3 - 6 in Figure \ref{fig:SHH_phase_time}. Plots show the bulk motion of the in-falling fluid converted into internal energy as the material converges and pressure begins to build.

\section{Discussion and Conclusions}
\label{sec:conclude}
The numerical and physical limitations of traditional SIDM numerical methods in the high-density limit of the collapse process were key motivating factors for this work. Semi-analytic methods, while more efficient at reaching and resolving greater core densities, predicate spherical symmetry and cannot readily accommodate the complex assembly history of massive halos. Existing numerical SIDM scattering algorithms do have the non-homogeneous and dynamic flexibility of galaxy simulations, but the Ansatz used to motivate the technique makes assumptions that break down in the frequent scattering limit. 

Our proposed SIDM-hydro hybrid (SHH) method takes a step to address these challenges by coupling the SIDM scattering algorithm to SPH gas particles via an artificial bridging function. The function uses density as a proxy for the local Knudsen number to smoothly interpolate between the kinetic and fluid solutions. In the low-scattering limit, the familiar SIDM algorithm is recovered, while at sufficiently high densities the simulation particles follow the Euler fluid equations. Allowing the macroscopic description of SIDM to transition to a fluid description integrates collapse-relevant physics, such as pressure and internal energy, into the solution. On the computational front, incorporation of hydrodynamic force terms alleviates the time-step crunch encountered in frequent-scattering SIDM environments. 

Although the resulting halo evolution shares basic features expected from other techniques, the physics captured by this first step presented here -- interpolating the momentum and energy equations only in the ideal-fluid limit -- is different from that implied by other treatments. In particular, the dynamics used here do not impose hydrostatic equilibrium, and do not include the non-ideal conduction terms, in marked contrast to the gravothermal fluid equations. The SHH technique can and should be extended, but the details of how collapse proceeds in the simplest SHH formulation presented here provide an illuminating contrast to other classic \citep{Burkert:2000,Yoshida:2000,Balberg_2002,Koda:2011} and recent \citep{kamionkowski:2025,gurian:2025} treatments. 

We implemented the SHH scheme in the multi-method physics code GIZMO and used it to simulate the gravothermal collapse of $1.14 \cdot 10^9$ solar mass halos for a 50 $\mathrm{cm}^2/\mathrm{g}$ SIDM particle. We verified that the method converges with particle resolution, and compared the impact of simulation parameter choices on collapse time, relative to equivalent SIDM-only halos. Previous works \citep{Palubski:2024ibb, Mace:2024uze, Fischer:2024eaz} have pointed out how sensitive SIDM modeling is to these choices. We found that choices of $\kappa$ and particle resolution play a less significant role in SHH simulations than in their SIDM-only counterparts. This is because SHH transitions to the fluid solution in the frequent-scattering limit, where the choice of simulation parameters and resolution is particularly pertinent. 

Previous works also found that adaptive force softening is unappealing for SIDM methods as it may hasten core collapse via numerical heating. At the same time, it is also undesirable to apply different softening methods to hydro and gravity terms which can result in inconsistent solutions. We here opt for the first of the two evils, and prioritize the self-consistency of the gravitational potential as the dense core forms. In fact, the presence of the hydro terms appears to reduce some of the numerical heating effects while also reducing computation time. We are encouraged to observe that despite the use of adaptive softening, we encounter no core density reversal, and little to no stalling in our hybrid method. Nonetheless, the effect of force softening should be investigated in greater depth, and we leave this for future work. We find that computational expense and energy conservation also compares favorably with existing SIDM methods (see Figure \ref{fig:SHH_SIDM_timebin}). 

In both our SIDM-only and hybrid method simulations we reproduce the qualitative behavior shown in semi-analytic and numerical simulation up to the early stages of core collapse. We find that the initial broadening of the core and its associated reduction in density is less acute than in some of our reference models, however as this appears in equal parts for both methods, this is likely due to difference in initial conditions and configuration choices. Depending on the choice of bridging function, we are able to follow core densities up to $10^4$ - $10^5\rho_{c}$, which is an order of magnitude improvement from the SIDM-only halos. We also find evidence for the formation of a hot core for which gravitational and hydro forces appear to balance each other, as the fluid becomes pressure supported. 

The bridging function plays a significant role in setting both the collapse time and the final density and geometry of the core. While the transition into the fluid regime at $\mathrm{Kn} \approx 10^{-3}$ is relatively well defined, there is no equivalent condition to demark its kinetic counterpart. We selected bridging functions with hydro force contributions at $\mathrm{Kn} = 1$ ranging from $2\cdot 10^{-2}$ to $10^{-4}$, in conjunction with gradients simulating gradual to sharp onset of fluid behavior. Halos with greater hydro contributions in low-density regions, which are important during the early evolution of the halo, tend to drive collapse to earlier times (both due to the contribution of the hydro term to the motion, and reduction of numerical energy loss due to SIDM scattering). However, at later times, when the dense core has already begun to form, the sharpness with which the hydro term comes into effect also plays a role, as well as how close $h(\rho)$ is to unity once $\mathrm{Kn} = 10^{-3}$.

While the aforementioned results are promising, there remain clear short-comings both from a numerical and physical perspective. Physically, the gravothermal collapse is generally understood to be a quasi-hydrostatic process, because the timescale for energy transport due to scatterings is typically long compared to the dynamical timescale of the halo. In the SHH method, the system can deviate from hydrostatic equilibrium during the intermediate stages of the collapse. Numerically, perhaps most pertinent is the choice of bridging function, which, like the choice of $\beta$ in the gravothermal semi-analytic models, is not derived from first principles. To a degree this is the trade off for side-stepping solving the Boltzmann equation directly, however a number of steps can feasibly be taken to constrain the bridging function and connect its parameters to the physical properties of the fluid and dark matter particle in a more direct manner. Alternatively, more efficient Boltzmann solvers, as for example proposed by \cite{gurian:2025}, could be used over that interval though the issue of domain matching would have to be addressed.

Shrinking the interval over which the two methods are interpolated would be an obvious way to address the problem without resorting to an additional fluid solver. For the present our hybrid method takes $\mathrm{Kn} = 10^{-3}$ as the transition to the ideal fluid domain. The non-ideal fluid terms were shown in Eq.~(\ref{eq:NavierStokes}), which are derived as an expansion around the equilibrium Maxwell-Boltzmann solution of an ideal fluid, where the Knudsen number is generally used as the expansion parameter. Including these terms would push the transition into the fluid treatment to appreciably larger Knudsen numbers. In addition, the inclusion of thermal conductivity and viscosity would add important mechanisms via which energy transported within a SIDM halo. These were not captured by the implementation of the method we considered here. (GIZMO already supports non-ideal fluid terms for its SPH module, though additional numerical caveats must be taken into account that arise from the Lagrangian form of these terms.)

We made a number of simplifying assumptions in our implementation of the bridging function for expository purposes. In particular, we used the local density as a proxy for the cell Knudsen number via a simple geometric relationship (Eq.~\ref{equ:Knudsen_sims}), and then used this to parameterize the bridging function. However, both the physical and cell Knudsen number could be taken directly from the force softening $h_{s}$ and various other physical quantities and gradients already tracked by numerical fluid dynamic solvers. This would also be advantageous in scenarios where the dark matter is made up of more than one particle and there are multiple scattering cross-sections depending on the chemical make-up of the fluid. 

On the numerical front, SIDM retains its well documented idiosyncrasies such as non-reversibility of timestep changes and broken time symmetry of the scattering kick. These features remain in SHH when the hydro component is strongly suppressed at lower densities. Extending the hydro solution into the non-ideal regime would also alleviate some of these numerical problems as it would allow a transition into the fluid solution at an earlier stage of collapse. The SPH solution of course comes with its own well-documented limitations, especially upon inclusion of non-ideal fluid terms. By construction, the Lagrangian formulation of the Euler equation introduces low-order inconsistencies, contact discontinuities, and excessive diffusion due to the artificial viscosity term introduced to resolve shocks. These can be addressed via a range of numerical techniques though often at a computational cost or introduction of another type of error. Increasing the nearest-neighbor particles of the SPH kernel to $\thicksim 100$  will decrease these errors though as was demonstrated in previous SIDM studies, the method is very sensitive to configuration parameters, and broadening the scattering kernel could artificially hasten collapse. Even outside the hybrid scenario, the correct numerical implementation of SIDM dynamics remains incredibly important, and these nuances should be kept at the forefront of developing new methods.

We have restricted considerations in this work to the simplest self-interacting dark matter model, though of course much richer, well-motivated dark sector phenomenology has been proposed \citep{Ackerman2009, Agrawal_2017,Boddy2016}. Extending existing numerical SIDM methods also serves to accommodate more of these additional dark-matter effects. Imbuing the SIDM particles with gas properties creates a natural way to use existing hydro-fluid infrastructure to include dark cooling \citep{Ryan_2022a, Ryan_2022b}. This simple extension to the dark sector could have profound consequences for the further evolution of the collapsed SIDM cores, like formation of exotic dense objects such as black holes below the Chandrasekhar limit, or a new channel to produce super massive black hole seeds. Having a fully three-dimensional simulation, able to trace collapse well into the frequent scattering regime is vital to exploring these possibilities.  

Lastly, all the halos in this work are treated as isolated objects, with no baryonic content, in which the SIDM scattering is "turned on" and the fully formed NFW profiles is taken as the initial condition. Incorporation of SIDM effects through collapse, as well in mergers and accretion in either the growth or disruption of the gravothermally collapsed cores would be of great interest. Three-dimensional simulations, non-reliant on spherical symmetry, are well suited to build self-consistent SIDM halo assembly models that take environmental factors into account. An obvious extension is the inclusion of baryonic physics ranging from the presence of cooling gas, to star formation, and even stellar feedback, to ascertain the respective impact dark and Standard Model fluids have on each other over the lifetime of a galactic halo.    

\section*{Acknowledgments}
 We thank James Gurian for collaboration on early stages of this work and for very useful discussions on several aspects of the final results, especially the role of hydrostatic equilibrium. We thank Kim Boddy for helpful discussions on the standard SIDM methods. This work was supported at Pennsylvania State University by NASA ATP Program No. 80NSSC22K081. S. Schon was partially supported by the Eberly College of Science. D.J.~was supported by NSF grants (AST-2307026, AST-2407298) at Pennsylvania State University and by KIAS Individual Grant PG088301 at Korea Institute for Advanced Study.

This work made use of the following software packages: \texttt{GIZMO} \citep{Hopkins2015, Springel2005}, \texttt{matplotlib} \citep{Hunter:2007}, \texttt{numpy} \citep{numpy}, \texttt{python} \citep{python}, \texttt{Cython} \citep{cython:2011}, \texttt{h5py} \citep{collette_python_hdf5_2014, h5py_7560547}, and \texttt{tqdm} \citep{tqdm_2560836}. Software citation information aggregated using \texttt{\href{https://www.tomwagg.com/software-citation-station/}{The Software Citation Station}} \citep{software-citation-station-paper, software-citation-station-zenodo}.

\section*{Data Availability}
The modified version of GIZMO, as well as snapshots from the simulations, are available from the authors upon request. 

\bibliographystyle{mnras}
\bibliography{References} 

\bsp	
\label{lastpage}
\end{document}